\documentclass{aa}  

\usepackage{graphicx}
\usepackage{txfonts,textcomp}
\usepackage{gensymb} 
\usepackage{placeins}
\usepackage{natbib,twoopt}
\usepackage[breaklinks=true]{hyperref} 
\bibpunct{(}{)}{;}{a}{}{,} 
\makeatletter
\newcommandtwoopt{\citeads}[3][][]{\href{http://adsabs.harvard.edu/abs/#3}%
{\def\hyper@linkstart##1##2{}%
\let\hyper@linkend\@empty\citealp[#1][#2]{#3}}}
\newcommandtwoopt{\citepads}[3][][]{\href{http://adsabs.harvard.edu/abs/#3}%
{\def\hyper@linkstart##1##2{}
\let\hyper@linkend\@empty\citep[#1][#2]{#3}}}
\newcommandtwoopt{\citetads}[3][][]{\href{http://adsabs.harvard.edu/abs/#3}%
{\def\hyper@linkstart##1##2{}
\let\hyper@linkend\@empty\citet[#1][#2]{#3}}}
\newcommandtwoopt{\citeyearads}[3][][]%
{\href{http://adsabs.harvard.edu/abs/#3}
{\def\hyper@linkstart##1##2{}%
\let\hyper@linkend\@empty\citeyear[#1][#2]{#3}}}
\makeatother
\def\m2s2{\hbox{\,m$^{2}$\,s$^{-2}$}} 
\def\Msun{$M_{\odot}$\xspace}             
\def\Rsun{$R_{\odot}$\xspace}
\def\Mjup{\hbox{$\mathrm{M}_{\rm J}$}}
\def\Rjup{\hbox{$\mathrm{R}_{\rm J}$}}

\def\ten[#1]{$\;\times 10^{#1}$}

\newcommand{\Rnom}{\hbox{$\mathcal{R}^{\rm N}_{\odot}$}} 

\newcommand{\Renom}{\hbox{$\mathcal{R}^{\rm N}_{e \rm E}$}}

\newcommand{\RJnom}{\hbox{$\mathcal{R}^{\rm N}_{e \rm J}$}}

\newcommand{\rebound}{{\sc \tt REBOUND}\xspace}
\newcommand{\whf}{{\sc \tt WHFast}\xspace}
\newcommand{\emcee}{{\sc \tt emcee}\xspace}
\newcommand{\juliet}{{\sc \tt juliet}\xspace}
\newcommand{\batman}{{\sc \tt batman}\xspace}

\newcommand{\celerite}{{\sc \tt celerite}\xspace}

\newcommand{\dynesty}{{\sc \tt dynesty}\xspace}

\newcommand{\optimize}{{\sc \tt optimize}\xspace}
\newcommand{\scipy}{{\sc \tt scipy}\xspace}

\newcommand{\MEarth}{$\mathrm{M_E}$\xspace}

\def\Msun{$M_{\odot}$\xspace}            
\def\Rsun{$R_{\odot}$\xspace}
\newcommand{\orcid}[1]{\protect\href{https://orcid.org/#1}{\protect\includegraphics[width=8pt]{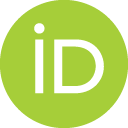}}}
\hypersetup{colorlinks=true, citecolor=blue, linkcolor=blue}

\begin{document} 

   \title{Evidence for transit-timing variations of the 11~Myr exoplanet TOI-1227\,b}

   \author{
        J.M.~Almenara\orcid{0000-0003-3208-9815}\inst{\ref{geneva},\ref{grenoble}}
        \and X.~Bonfils\orcid{0000-0001-9003-8894}\inst{\ref{grenoble}}
        \and T.~Guillot\orcid{0000-0002-7188-8428}\inst{\ref{nice}}
        \and M.~Timmermans\inst{\ref{liege}}
        \and R.F.~D\'{i}az\inst{\ref{ba}}
        \and J.~Venturini\orcid{0000-0001-9527-2903}\inst{\ref{geneva}}
        \and A.C.~Petit\orcid{0000-0003-1970-1790}\inst{\ref{nice}}
        \and T.~Forveille\orcid{0000-0003-0536-4607}\inst{\ref{grenoble}}
        \and O.~Su\'{a}rez\inst{\ref{nice}}
        \and D.~M\'{e}karnia\inst{\ref{nice}}
        \and A.H.M.J.~Triaud\orcid{0000-0002-5510-8751}\inst{\ref{birmingham}}
        \and L.~Abe\inst{\ref{nice}}
        \and P.~Bendjoya\inst{\ref{nice}}
        \and F.~Bouchy\orcid{0000-0002-7613-393X}\inst{\ref{geneva}}
        \and J.~Bouvier\inst{\ref{grenoble}}
        \and L.~Delrez \inst{\ref{liege}}
        \and G.~Dransfield\inst{\ref{birmingham}}
        \and E.~Ducrot\inst{\ref{cea}}
        \and M.~Gillon \inst{\ref{liege}}
        \and M.J.~Hooton\orcid{0000-0003-0030-332X} \inst{\ref{cambridge}}
        \and E.~Jehin\orcid{0000-0001-8923-488X}\inst{\ref{STAR}}
        \and A.W.~Mann\orcid{0000-0003-3654-1602}\inst{\ref{UNC}}
        \and R.~Mardling\inst{\ref{australia},\ref{geneva}}
        \and F.~Murgas\orcid{0000-0001-9087-1245}\inst{\ref{iac},\ref{ull}}
        \and A.~Leleu\inst{\ref{geneva}}
        \and M.~Lendl\orcid{0000-0001-9699-1459}\inst{\ref{geneva}}
        \and D.~Queloz \inst{\ref{cambridge}}
        \and S.~Seager\orcid{0000-0002-6892-6948}\inst{\ref{mit},\ref{mit2},\ref{mit3}}
        \and J.~Winn\inst{\ref{princeton}}
        \and S.~Zu$\rm \Tilde{n}$iga Fern\'andez\orcid{0000-0002-9350-830X} \inst{\ref{liege}}
    }

      \institute{
        Observatoire de Gen\`eve, Département d’Astronomie, Universit\'e de Gen\`eve, Chemin Pegasi 51b, 1290 Versoix, Switzerland\label{geneva}
        \and Univ. Grenoble Alpes, CNRS, IPAG, F-38000 Grenoble, France\label{grenoble}
        \and Universit\'e C\^ote d'Azur, Laboratoire Lagrange, OCA, CNRS UMR 7293, Nice, France\label{nice}
        \and Astrobiology Research Unit, Universit\'e de Li\`ege, All\'ee du 6 Ao\^ut 19C, B-4000 Li\`ege, Belgium\label{liege}
        \and International Center for Advanced Studies (ICAS) and ICIFI (CONICET), ECyT-UNSAM, Campus Miguelete, 25 de Mayo y Francia, (1650) Buenos Aires, Argentina\label{ba}
        \and School of Physics \& Astronomy, University of Birmingham, Edgbaston, Birmingham B15 2TT, UK\label{birmingham}
        \and AIM, CEA, CNRS, Universit\'e Paris-Saclay, Universit\'e de Paris, F-91191 Gif-sur-Yvette, France\label{cea}
        \and Cavendish Laboratory, JJ Thomson Avenue, Cambridge CB3 0HE, UK\label{cambridge}
        \and Space Sciences, Technologies and Astrophysics Research (STAR) Institute, Universit\'e de Li\`ege, All\'ee du 6 Ao\^ut 19C, B-4000 Li\`ege, Belgium\label{STAR}
        \and Department of Physics and Astronomy, The University of North Carolina at Chapel Hill, Chapel Hill, NC 27599, USA\label{UNC}
        \and School of Physics \& Astronomy, Monash University, Victoria, 3800, Australia\label{australia}
        \and Instituto de Astrofísica de Canarias (IAC), E-38200 La Laguna, Tenerife, Spain\label{iac}
        \and Dept. Astrofísica, Universidad de La Laguna (ULL), E-38206 La Laguna, Tenerife, Spain\label{ull}
        \and Department of Physics and Kavli Institute for Astrophysics and Space Research, Massachusetts Institute of Technology, Cambridge, MA 02139, USA\label{mit}
        \and Department of Earth, Atmospheric and Planetary Sciences, Massachusetts Institute of Technology, Cambridge, MA 02139, USA\label{mit2}
        \and Department of Aeronautics and Astronautics, MIT, 77 Massachusetts Avenue, Cambridge, MA 02139, USA\label{mit3}
        \and Department of Astrophysical Sciences, Princeton University, NJ 08544, USA\label{princeton}
    }   

\date{Received ; Accepted}

 
  \abstract
  {
  TOI-1227\,b is an 11~Myr old validated transiting planet in the middle of its contraction phase, with a current radius of 0.85~\Rjup. It orbits a low-mass pre-main sequence star (0.170~\Msun, 0.56~\Rsun) every 27.4~days. The magnetic activity of its young host star induces radial velocity jitter and prevents good measurements of the planetary mass. We gathered additional transit observations of TOI-1227\,b with space- and ground-based telescopes, and we detected highly significant transit-timing variations (TTVs). Their amplitude is about 40~minutes and their dominant timescale is longer than 3.7~years. Their most probable origin is dynamical interactions with additional planets in the system. We modeled the TTVs with inner and outer perturbers near first and second order resonances; several orbital configurations provide an acceptable fit. More data are needed to determine the actual orbital configuration and eventually measure the planetary masses. These TTVs and an updated transit chromaticity analysis reinforce the evidence that TOI-1227\,b is a planet.
  }


   \maketitle
%

\section{Introduction}

Young planets offer unique insight into the early physical processes that shape planetary systems, such as the post-formation cooling and contraction of the planets \citep{Vazan13, Mordasini12, Linder19}, their orbital evolution \citep{Matsumura10, Bolmont17}, and the influence of stellar radiation on planetary evolution \citep{Lopez12, Owen17, Kubyshkina20}. In particular, very young resonant multi-planetary systems allow the resonance configuration to be observed as it formed in the protoplanetary disk \citep{Teyssandier2020}, before tidal dissipation erases the pristine configuration \citep{Papaloizou2010,Delisle2014}.

The planets of AU~Mic \citep{plavchan2020,cale2021} and V1298\,Tau\,b and e \citep{david2019,suarez2021,finociety2023} unfortunately remain the only planets with both mass and radius measurements for ages below 50~Myr. TOI-1227\,b \citep{mann2022} is, alongside K2-33 \citep{david2016,mann2016}, the youngest transiting planet known to date, with an age of just $11\pm2$~Myr \citep[estimated due to its membership to the Musca group,][]{mann2022}. The planet orbits a very low-mass pre-main sequence star ($0.170\pm0.015$~\Msun, $0.56\pm0.03$~\Rsun) every 27.4~days, which makes TOI-1227\,b the youngest transiting planet around the least massive star known today.

Young planets are key for theory of planet formation and evolution since they provide direct observational constraints just at the interface of the two stages. Such constraints will be crucial, for instance, to shed light on the current debate about the origin and composition of super-Earths and mini-Neptunes, where some studies propose a solely post-formation atmospheric-loss explanation \citep{Owen17}, while others highlight the importance of planet formation in producing the two populations of planets \citep{Venturini20}. In addition, different evolutionary processes such as photoevaporation \citep{Owen17} and core-powered mass loss \citep{Gupta19} yield equal outputs after gigayears of evolution, but they act on different timescales when scrutinizing at the $\sim$10-100 Myr level.
Furthermore, several physical parameters affect the size of a planet over time, such as the core and envelope mass and composition \citep{Mordasini20, Dorn_2017}, the presence of compositional gradients \citep{Vazan13}, internal luminosity \citep{Guillot10}, the extension of convective and radiative regions \citep{Haldemann23}, and the extreme-ultraviolet (XUV) environment \citep{Kubyshkina20}; however, their distinctive imprint fades for mature, gigayear-old planets. On the contrary, the size of young planets is much more sensitive to these model assumptions \citep{Mordasini12, Vazan13, Muller20}, which suggest the possibility of disentangling among them when observing planets that just emerged from the protoplanetary disk.

TOI-1227\,b has a measured size of 0.85~\Rjup, and its mass is not known. Planet evolution models indicate that in order to match the planet size at its inferred age, the planet mass could range between approximately 5 and 50~\MEarth, depending on different model assumptions \citep{Mordasini12, Linder19, mann2022}.
The planet is believed to be halfway through its contraction phase, and it is expected to mature into a planet of roughly Neptune size when reaching an age of approximately a gigayear \citep{mann2022}. 
Determining the planet's mass is key to set proper constraints on the model assumptions mentioned above (otherwise the unknown mass becomes a confounding factor). Radial velocity measurements are very challenging for this target due to the stellar activity. The starspots on fast-rotating young stars do indeed induce large apparent radial velocity variations \citep[e.g.,][]{donati2023}, blurring any planetary signal. Because of this, only an upper mass limit of $\simeq$0.5~\Mjup\ could be reported from past radial velocity campaigns \citep{mann2022}. Due to the difficulties entailed in obtaining radial velocities for a target of these characteristics, transit-timing variations \citep[TTVs,][]{agol2005,holman2005} from dynamical interactions offer an alternative and promising path to constrain the planet mass, not only for TOI-1227\,b, but in general for any young multi-planetary system \citep{martioli2021}.

We observed additional transits of TOI-1227\,b and detected TTVs, which suggest that the system contains at least one additional planet. The article is organized as follows: We present the new transit photometry of TOI-1227\,b in Sect.~\ref{section.observations}, we explain how we derived the transit timings in Sect.~\ref{section.juliet} and modeled them in Sect.~\ref{section.model}. Finally, in Sect.~\ref{section.results} we discuss the results of our work.

\section{Observations}\label{section.observations}

We first collected the Transiting Exoplanet Survey Satellite \citep[TESS,][]{ricker2015} and ground-based transit photometry observations presented in \citet{mann2022}. To those literature data, we added new photometry gathered since by TESS, as well as follow-up ground-based observations with Antarctica Search for Transiting ExoPlanets (ASTEP), Exoplanets in Transits and their Atmospheres (ExTrA), Search for habitable Planets EClipsing ULtra-cOOl Stars (SPECULOOS-South), and TRAnsiting Planets and PlanetesImals Small Telescope (TRAPPIST-South)\footnote{The ground-based transit observations are available at \url{https://zenodo.org/records/10405623}}. Table~\ref{table:log} summarizes all transit observations used in this work, and we detail the new observations by telescope below. 

\begin{table*}
\caption{Log of transit observations.}             
\label{table:log}      
\centering                          
\begin{tabular}{r c c c c c c}        
\hline\hline                 
Epoch & Mid-transit date & Telescope & Band & Exposure time & Coverage & Source \\    
 & YYYY-MM-DD (UT) & & & [s] & & \\
\hline                        
   0  & 2019-05-13 & TESS s11 & TESS & 120 & Full & \citet{mann2022} \\      
   1  & 2019-06-10 & TESS s12 & TESS & 120 & Full & \citet{mann2022}  \\
   9  & 2020-01-15 & SOAR & i' & 120 & Full & \citet{mann2022}  \\
   13 & 2020-05-03 & LCOGT-SSO & i' & 200 & Partial & \citet{mann2022}  \\ 
   16 & 2020-07-24 & LCOGT-SAAO & r', z$_{\rm S}$ & 200 & Partial & \citet{mann2022} \\ 
   25 & 2021-03-28 & SOAR & g' & 120 & Partial & \citet{mann2022}  \\ 
   26 & 2021-04-24 & LCOGT-SSO & z$_{\rm S}$ & 210 & Partial & \citet{mann2022}  \\    
   26 & 2021-04-24 & ASTEP & ASTEP & 200 & Partial & \citet{mann2022}  \\ 
   27 & 2021-05-21 & TESS s38 & TESS & 120 & Full & \citet{mann2022}  \\ 
   28 & 2021-06-18 & LCOGT-CTIO & g' & 300 & Partial & \citet{mann2022}  \\ 
   28 & 2021-06-18 & ExTrA T23 & ExTrA & 60 & Partial & This work\\ 
   39 & 2022-04-15 & TRAPPIST-South & \textit{I+z} & 180 & Full & This work \\ 
   39 & 2022-04-15 & ExTrA T23 & ExTrA & 60 & Partial & This work \\ 
   40 & 2022-05-12 & ExTrA T3 & ExTrA & 60 & Partial & This work \\ 
   51 & 2023-03-09 & ExTrA T123 & ExTrA & 60 & Partial & This work \\ 
   53 & 2023-05-03 & TESS s64 & TESS & 120 & Full & This work \\ 
   53 & 2023-05-03 & SPECULOOS-Io & z' & 21 & Full & This work \\ 
   53 & 2023-05-03 & ExTrA T123 & ExTrA & 60 & Full & This work \\ 
   54 & 2023-05-30 & TESS s65 & TESS & 120 & Full & This work \\ 
   54 & 2023-05-30 & ASTEP & ASTEP+ Red & 200 & Full & This work \\ 
   55 & 2023-06-26 & ASTEP & ASTEP+ Red & 200 & Full & This work \\ 
   57 & 2023-08-20 & ASTEP & ASTEP+ Red & 200 & Partial & This work \\ 
\hline                                   
\end{tabular}
\tablefoot{The epoch is the number of orbital periods since the first observed transit. For the TESS observations, the sector number is specified after "s." For the ExTrA observations, "T" stands for "telescope" and the numbers following it represent the specific telescopes that observed the transit. For example, "T123" means that telescopes 1, 2, and 3 all observed the transit.}
\end{table*}

\subsection{ASTEP}
The ASTEP is a 0.4-m telescope equipped with a Wynne Newtonian coma corrector, located at Dome C on the east Antarctic plateau \citep{Guillot2015,Mekarnia2016}. Until December 2021, it was equipped with a 4k × 4k front-illuminated FLI Proline KAF-16801E CCD with an image scale of $0.93\arcsec \rm{pixel}^{-1}$ resulting in a $1^{\circ} \times 1^{\circ}$ corrected field of view. The focal instrument dichroic plate split the beam into a blue wavelength channel for guiding, and a non-filtered red science channel roughly matching an $Rc$ transmission curve \citep{Abe2013}.
In January 2022, the focal box was replaced with a new one with two high sensitivity cameras including an Andor iKon-L936 at red wavelengths. The image scale is $1.39\arcsec \rm{pixel}^{-1}$ with a transmission curve centered on 850$\pm$138~nm \citep{Schmider2022}. The telescope is automated or remotely operated when needed. Due to the extremely low data transmission rate at the Concordia station, the data are processed on-site using IDL \citep{Mekarnia2016} and Python \citep{Dransfield2022} aperture photometry pipelines. The raw light curves of
up to 1\,000 stars in the field are transferred to Europe on a server in Rome, Italy, and they are then available for deeper analysis.

\begin{figure}[htbp]
  \includegraphics[width=0.45\textwidth]{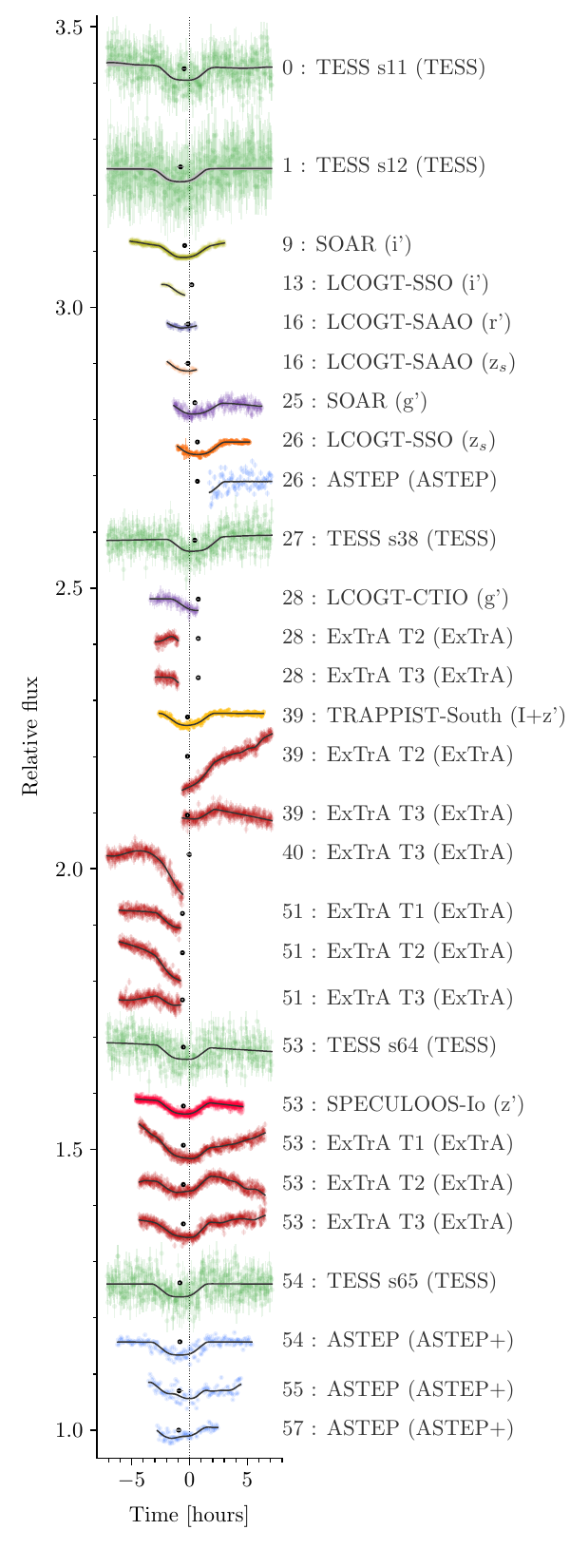}\vspace{-0.5cm}
  \caption{Transit observations (vertical error bars) with the model that combines both transit and noise (median as a black line and 68\% interval in gray), arbitrarily shifted vertically. Each transit observation is labeled with the transit epoch, the telescope, and the observing band. Each transit is centered relative to the maximum a posteriori value of the linear ephemeris derived in Sect.~\ref{section.sine}. Horizontal error bars show the mid-transit time offset from the linear ephemeris.} \label{fig.phot}
\end{figure}

Four observations of TOI-1227\,b were carried out with ASTEP, all under clear sky conditions with winds between 2 and 5~ms$^{-1}$ and a temperature ranging between $-$65 and $-$70$^{\circ}$C. An egress was observed on 24 April 2021 with an image full width at half maximum (FWHM) of $4.9''$, and an almost complete transit, with the beginning of the transit missing, and on 20 August 2023 with a FWHM of $5.0''$. Full transits were observed on 30 May 2023 and 26 June 2023 with a FWHM of $6.8''$ and $6.0''$, respectively. The Moon was present for the 24 April 2021 (90\% illuminated), 30 May 2023 (76\% illuminated), and 26 June 2023 (51\% illuminated) observations. The light curves are shown in Fig.~\ref{fig.phot}.

\subsection{ExTrA}
The ExTrA \citep{Bon2015} is a low-resolution near-infrared (0.85 to 1.55~$\mu$m) multi-object spectrograph fed by three 60-cm telescopes located at La Silla Observatory in Chile. One full and four partial transits were observed using one, two, or three of the ExTrA telescopes. We used 8$\arcsec$ aperture fibers and the lowest-resolution mode ($R$$\sim$20) of the spectrograph, a combination that is optimal for the target's magnitude, with an exposure time of 60~seconds. Five fibers are positioned in the focal plane of each telescope to select light from the target and four comparison stars\footnote{2MASS J12235796-7230315, 2MASS J12280146-7208138, 2MASS J12243653-7207571, and 2MASS J12214622-7221353.}. We chose comparison stars with 2MASS $J$ magnitudes \citep{Skrutskie2006} and effective temperatures \citep{gaia2018} similar to the target. The resulting ExTrA data were analyzed using custom data reduction software.

\subsection{SPECULOOS}
The SPECULOOS Southern Observatory is located at ESO Paranal Observatory in Chile \citep{Jehin2018Msngr,Delrez2018,Sebastian_2021AA}. It holds four Ritchey-Chr\'etien 1m-class telescopes, each equipped with a deep-depletion Andor $2048 \times 2048$ CCD camera with a pixel scale of $0.35\arcsec$, which gives a field of view of $12 \times 12 \arcmin$. We observed one full transit of TOI-1227\,b in the \textit{Sloan-z'} filter on the night of 3 May 2023 with an exposure of 21 seconds, totaling 1033 measurements. The data reduction and photometry extraction were performed with a custom pipeline built with the \texttt{prose} package \citep{garcia2022}.

\subsection{TESS}
In addition to the sectors 11, 12, and 38 TESS data presented in \citet{mann2022}, TESS observed two full transits of TOI-1227\,b in sectors 64 and 65 with 2-min cadence. For the analysis in Sect.~\ref{section.juliet}, we used the presearch data conditioning simple aperture photometry (PDCSAP; \citealt{smith2012}, \citealt{stumpe2012,stumpe2014}, \citealt{caldwell2020}) light curve of TOI-1227, produced by the TESS Science Processing Operations Center \citep[SPOC;][]{jenkins2016,caldwell2020}.

\subsection{TRAPPIST}
The TRAPPIST-South is a 0.6-m Ritchey-Chr\'etien telescope located at ESO La Silla Observatory in Chile \citep{Jehin2011,Gillon2011}. It is equipped with a German equatorial mount and has a $2048 \times 2048$ back-illuminated FLI ProLine PL3041-BB CCD camera with a pixel scale of $0.64\arcsec$, giving a field of view of $22 \times 22 \arcmin$. One full transit of TOI-1227\,b was observed on 15 April 2022 with an exposure time of 180 seconds in the \textit{I+z} custom filter, and we collected 170 measurements during the observation. We performed the data analysis using a custom pipeline built with the \texttt{prose} package, which allowed for the data to be reduced as well as for differential photometry to be performed.

\medskip
In the transit light curves presented in this work, no detrending was applied for systematics. Any systematics present in the light curves were addressed during the global analysis (Sect.~\ref{section.juliet}).

\section{Transit photometry analysis}\label{section.juliet}
We derived the transit times of TOI-1227\,b with \juliet \citep{espinoza2019}, using \batman \citep{kreidberg2015} for the transit model, and the approximate Matern kernel Gaussian process (GP) included in \celerite \citep{foreman-mackey2017} to model the systematics. We used different GPs' hyperparameters for each telescope and for each transit observation. The timing of each transit epoch, observed with one to five telescopes, is a free parameter. We assumed a circular orbit, but this assumption should not significantly affect the determination of the transit times.
We adopted non-informative or broad priors for all the parameters. To sample from the posterior, we used the \dynesty nested sampling code \citep{speagle2020}. The complete list of parameters, priors, and posteriors' median and 68.3\% credible interval (CI) are shown in Table~\ref{table.params}. We scaled the radius of TOI-1227\,b using the \citet{mann2022} stellar radius, $0.56\pm0.03$~\Rsun. Figure~\ref{fig.phot} shows the data and the model posterior. We explain how we briefly tested the accuracy of our transit timings using simulated transits in Sect.~\ref{sec.referee}.

\begin{table*}
  \small
  \setlength{\tabcolsep}{5pt}
\renewcommand{\arraystretch}{1.1}
\centering
\caption{Inferred parameters from the transit photometry analysis.}\label{table.params}
\begin{tabular}{lccc}
\hline
\hline
Parameter & Units & Prior & Posterior median and 68.3\% CI  \\
\hline
\emph{\bf TOI-1227} \\
Mean density, $\rho_{\star}$     & [$\mathrm{g\;cm^{-3}}$] & $U(0.9, 1.6)$ & $1.197 \pm 0.063$ \\
$q_1$ ASTEP, ExTrA, TESS &                   & $U(0, 1)$ & $0.18^{+0.18}_{-0.12}$, $0.19^{+0.22}_{-0.13}$, $0.25^{+0.31}_{-0.17}$ \\
$q_2$ ASTEP, ExTrA, TESS &                   & $U(0, 1)$ & $0.62^{+0.27}_{-0.36}$, $0.33^{+0.37}_{-0.24}$, $0.32^{+0.37}_{-0.23}$ \\
$q_1$ g', r', i' &                     & $U(0, 1)$ & $0.70^{+0.20}_{-0.26}$, $0.50 \pm 0.34$, $0.77^{+0.15}_{-0.20}$ \\
$q_2$ g', r', i' &                     & $U(0, 1)$ & $0.71^{+0.20}_{-0.32}$, $0.63^{+0.27}_{-0.39}$, $0.64^{+0.21}_{-0.24}$ \\
$q_1$ z', z$_{\rm S}$, I+z &                     & $U(0, 1)$ & $0.71^{+0.19}_{-0.22}$, $0.61 \pm 0.24$, $0.883^{+0.086}_{-0.16}$ \\
$q_2$ z', z$_{\rm S}$, I+z &                     & $U(0, 1)$ & $0.26^{+0.24}_{-0.17}$, $0.43^{+0.29}_{-0.25}$, $0.46 \pm 0.21$ \smallskip \\

\emph{\bf TOI-1227\,b} \\
Semi-major axis, $a$                   & [au]             & & $0.0942 \pm 0.0053$ \\
Inclination, $i_p$                     & [\degree]        & & $88.711 \pm 0.036$ \\
Radius ratio, $R_{\mathrm{p}}/R_\star$ &                  & & $0.1567 \pm 0.0029$ \\
Scaled semi-major axis, $a/R_{\star}$  &                  & & $36.18 \pm 0.62$ \\
Impact parameter, $b$                  &                  & & $0.814 \pm 0.012$\\ 
Transit duration, $T_{14}$             & [h]              & & $4.752^{+0.052}_{-0.074}$ \\
Radius, $R_{\mathrm{p}}$               &[\Renom]           & & $9.58 \pm 0.55$ \\
                                       &[\RJnom]      & & $0.854 \pm 0.049$ \\
$r_1$                              &    & $U(0, 1)$ & $0.8759^{+0.0071}_{-0.0080}$ \\
$r_2$                              &    & $U(0.140, 0.175)$ & $0.1567 \pm 0.0029$ \smallskip \\

\emph{\bf Mid-transit times} \\
Epoch 0 (TESS s11) & [BJD$_{\mathrm{TDB}}$]   & $U(2458617.44, 2458617.53)$ & $2458617.4817 \pm 0.0081$ \\
Epoch 1 (TESS s12) & [BJD$_{\mathrm{TDB}}$]   & $U(2458644.76, 2458644.90)$ & $2458644.830 \pm 0.015$ \\
Epoch 9 (SOAR i') & [BJD$_{\mathrm{TDB}}$]   & $U(2458863.71, 2458863.76)$ & $2458863.7372 \pm 0.0028$ \\
Epoch 13 (LCOGT i') & [BJD$_{\mathrm{TDB}}$]   & $U(2458973.17, 2458973.25)$ & $2458973.2089 \pm 0.0048$ \\
Epoch 16 (LCOGT r', z$_{\rm S}$) & [BJD$_{\mathrm{TDB}}$]   & $U(2459055.230, 2459055.309)$ & $2459055.2799^{+0.0036}_{-0.012}$ \\
Epoch 25 (SOAR g') & [BJD$_{\mathrm{TDB}}$]   & $U(2459301.48, 2459301.65)$ & $2459301.5581^{+0.0062}_{-0.0052}$ \\
Epoch 26 (ASTEP, LCOGT z$_{\rm S}$) & [BJD$_{\mathrm{TDB}}$]   & $U(2459328.915, 2459328.945)$ & $2459328.9290 \pm 0.0015$ \\
Epoch 27 (TESS s38) & [BJD$_{\mathrm{TDB}}$]   & $U(2459356.24, 2459356.32)$ & $2459356.2810 \pm 0.0063$ \\
Epoch 28 (LCOGT g', ExTrA T23) & [BJD$_{\mathrm{TDB}}$]   & $U(2459383.60, 2459383.70)$ & $2459383.6550 \pm 0.0041$ \\
Epoch 39 (ExTrA T23, TRAPPIST-South I+z) & [BJD$_{\mathrm{TDB}}$]   & $U(2459684.58, 2459684.615)$ & $2459684.5931^{+0.0018}_{-0.0014}$ \\
Epoch 40 (ExTrA T3) & [BJD$_{\mathrm{TDB}}$]   & $U(2459711.900, 2459712.005)$ & $2459711.9615^{+0.0078}_{-0.010}$ \\
Epoch 51 (ExTrA T123) & [BJD$_{\mathrm{TDB}}$]   & $U(2460012.890, 2460012.935)$ & $2460012.9133 \pm 0.0029$ \\
Epoch 53 (ExTrA T123, SPECULOOS z', TESS s64) & [BJD$_{\mathrm{TDB}}$]   & $U(2460067.63, 2460067.65)$ & $2460067.6395 \pm 0.0012$ \\
Epoch 54 (ASTEP, TESS s65) & [BJD$_{\mathrm{TDB}}$]   & $U(2460094.97, 2460095.01)$ & $2460094.9886 \pm 0.0031$ \\
Epoch 55 (ASTEP) & [BJD$_{\mathrm{TDB}}$]   & $U(2460122.30, 2460122.40)$ & $2460122.3472^{+0.015}_{-0.0093}$ \\
Epoch 57 (ASTEP) & [BJD$_{\mathrm{TDB}}$]   & $U(2460177.02, 2460177.12)$ & $2460177.0688^{+0.0077}_{-0.0068}$ \smallskip\\

\emph{\bf Photometry} \\
Offset relative flux & [Relative flux]  & $N(0, 0.01)$ & $^{(a)}$ \\
Jitter               & [ppm]            & $J(1, 20000)$ & $^{(a)}$ \\
Amplitude of the GP  & [Relative flux]  & $J(10^{-6}, 1.0)$ & $^{(a)}$ \\
Timescale of the GP  & [days]           & $J(0.001, 10)$ & $^{(a)}$ \\
\hline
\end{tabular}
\tablefoot{Parameters without a prior are derived parameters. $^{(a)}$~The parameters listed for the photometry are different for each telescope and for each individual transit. The parameters $q_1$ and $q_2$ are the quadratic limb-darkening coefficients in the \citet{kipping2013} parametrization. The parameters $r_1$ and $r_2$ are the impact parameter and transit depth parametrized according to \citet{espinoza2018}. IAU 2012: \rm{au} = 149$\;$597$\;$870$\;$700~\rm{m}$\;$. IAU 2015: \Rnom = 6.957\ten[8]~\rm{m}, \Renom~=~6.378$\;$1\ten[6]~\rm{m}, and $\RJnom$ = 7.149$\;$2\ten[7]~\rm{m}. $U(a, b)$: A uniform distribution defined between a lower $a$ and an upper $b$ limit. $J(a, b)$: Jeffreys (or log-uniform) distribution defined between a lower $a$ and upper $b$ limit. $N(\mu, \sigma)$: Normal distribution prior with mean $\mu$, and standard deviation $\sigma$.}
\end{table*}

For comparison, we repeated the analysis for a model with equally spaced transits. This required using a linear ephemeris for the timing of each transit epoch ($E$), calculated as $T_0 + P \times E$, where $T_0$ is the transit time at a reference epoch and $P$ is the orbital period and both are free parameters. A Bayesian model comparison strongly favors \citep{kass1995} the model with TTVs over the linear ephemeris model, by a log-Bayes factor $\mathcal{Z}_{\rm TTVs}/\mathcal{Z}_{T_0,P}$ of $54.5\pm0.6$. This finding brings novel evidence that the transits of TOI-1227\,A are indeed produced by a planet, as eclipse-timing variations of the observed amplitude would imply a stellar mass perturber \citep{rappaport2013}. Such a perturber would have been observed in the spectra or the color-magnitude diagram \citep{mann2022}. The alternative false positive scenarios involving starspots or an accretion disk, presented in \citet{yu2015} for the young system PTFO\,8-8695, do not apply here. This is because the orbital period (27.4~days) is significantly different from the rotation period of the star (1.65~days) and TOI-1227 shows no evidence of having an accretion disk.

\begin{figure}[h]
   \centering
   \includegraphics[width=0.48\textwidth]{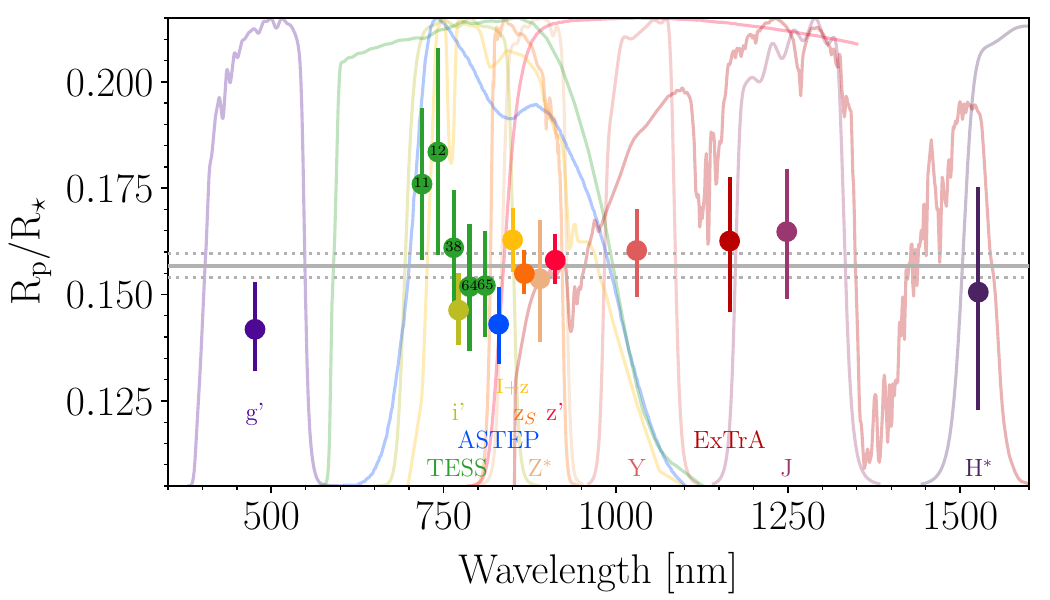}
      \caption{Posterior distribution comparison for $R_p/R_\star$ (error bars) computed for different bands (whose transmission curves are shown and labeled with different colors) and TESS sectors (shifted horizontally for visualization purposes, the TESS sector number is noted inside the dots). All transits were fitted jointly with a separate $R_p/R_\star$ parameter for each band. The horizontal solid and dotted gray lines represent the posterior median and 68.3\% CI assuming a common $R_p/R_\star$ (Sect.~\ref{section.juliet}).}
      \label{fig:RpRsRatio}
\end{figure}

We extended the transit chromatic analysis of \citet{mann2022} by including the additional photometric bands made available by our new observations. We excluded the epoch 16 transit (LCOGT-SAAO r', z$_{\rm S}$) which lacks a baseline and the partial transits observed by ExTrA. We separated the five TESS sectors because they use distinct photometric aperture and could therefore have different third light contamination. To refine the chromatic analysis, we synthesized photometry in the UKIRT-WFCAM filters\footnote{Retrieved from the SVO Filter Profile Service \citep[\url{http://svo2.cab.inta-csic.es/theory/fps/},][]{rodrigo2012,rodrigo2020}.} from the ExTrA spectrophotometry for the fully observed epoch 53 transit. This produced four intermediate-band light curves in addition to the light curve in ExTrA’s full wavelength range (0.85 to 1.55~$\mu$m), for a truncated Z band (Z$^{*}$), a Y, a J, and a truncated H band (H$^{*}$) (Fig.~\ref{fig.ExTrA_ZYJH}). This chromatic analysis is similar to the one presented above, except that the planet-to-star radius ratio ($R_p/R_\star$) has been adjusted separately for each dataset. This allowed for a different transit depth at each band, whose posteriors are listed in Table~\ref{table.RpRsRatio}. As Fig.~\ref{fig:RpRsRatio} shows, the transit depths for all of these bandpasses are consistent with a common $R_p/R_\star$. The observed achromaticity of the transit reinforces the hypothesis of a planet rather than a blended binary scenario.

\section{Modeling the transit timings}\label{section.model}
Although the TTVs are robustly detected, their modeling is challenging because the transit observations are heterogeneous and they do not cover the full range of the timing variability yet. Additionally, half of the epochs have only partial transit coverage, providing timings that are less robust than from a full transit observation. 

\subsection{Sine wave model}\label{section.sine}
\begin{figure}
  \centering
  \includegraphics[width=0.49\textwidth]{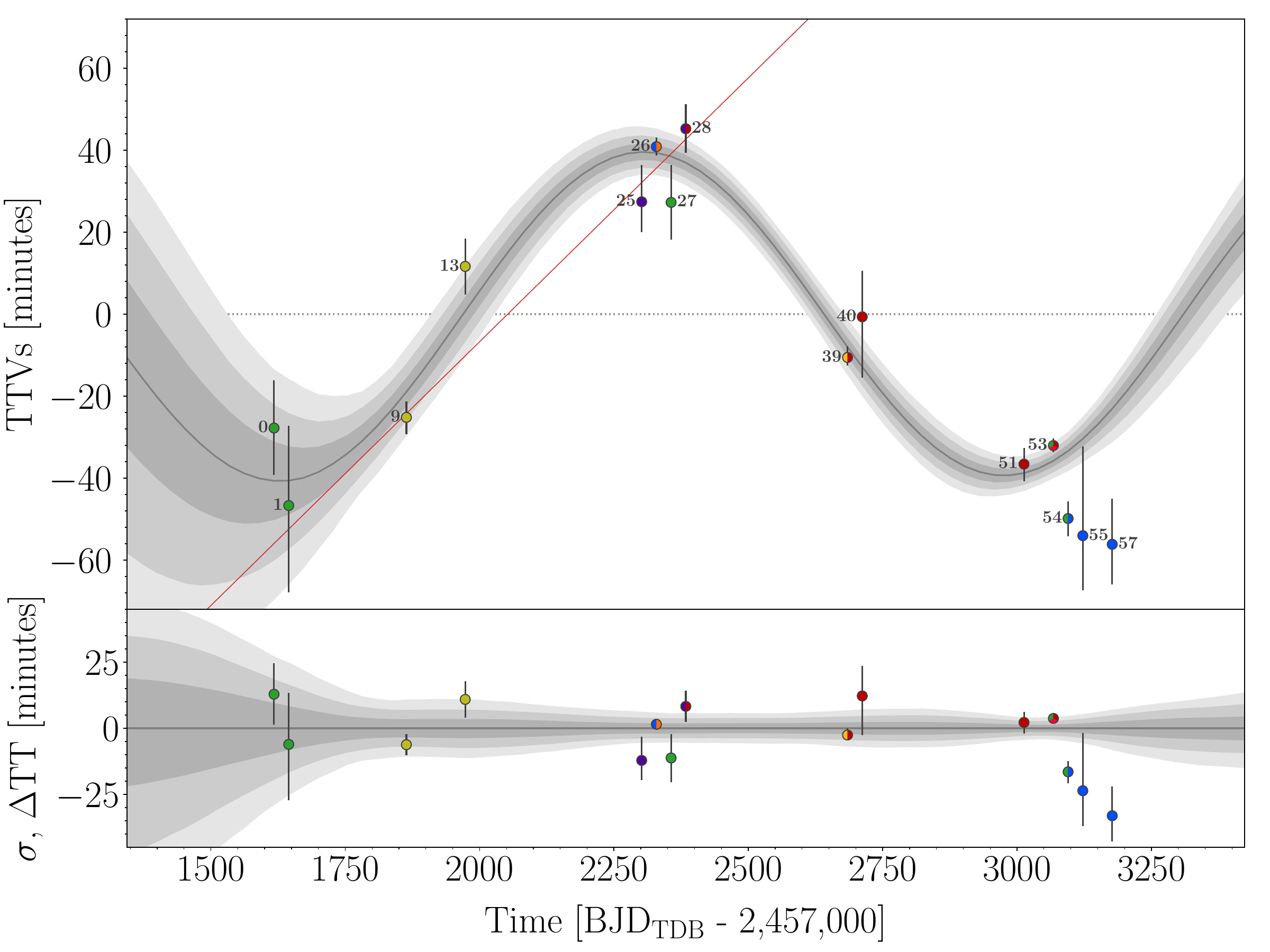}
  \caption{Posterior TTVs of TOI-1227\,b computed relative to a linear ephemeris (from the MAP value of the sinusoidal model) are shown with black error bars (Sect.~\ref{section.juliet}). In the upper panel, the TTVs' sinusoidal model (median, 1, 2, and 3-$\sigma$, in gray) is shown and compared with individual transit-time determinations (Sect.~\ref{section.juliet}, median and 68\% CI, the colors of the dots indicate the observing band as in Fig.~\ref{fig.phot}). The linear ephemeris from \citet{mann2022} is shown as a red line. In the lower panel, the posterior median of the TTVs' model was subtracted to visualize the uncertainty of the distribution and to allow for better comparison with the individual transit-time determinations.} \label{fig.TTVs}
\end{figure}

We started modeling the transit timings assuming a transit period that oscillates sinusoidally around an average value, which is often a good approximation for planets near first-order resonance \citep{lithwick2012} and inside resonance \citep{nesvorny2016}. We excluded from the analysis transit epoch 16, which covers neither the first nor the fourth contact. The posterior distribution was sampled using the \emcee\ algorithm \citep{goodmanweare2010, emcee}. We used uniform priors for all parameters. The posteriors are listed in Table~\ref{table.TTVsSine}, and Fig.~\ref{fig.TTVs} shows the observed TTVs with the sinusoidal model posterior. The maximum a posteriori (MAP) linear ephemeris, which was used to compute the TTVs in Figs.~\ref{fig.phot}, \ref{fig.TTVs}, \ref{fig.TTVsMMR}, and \ref{fig.model}, is BJD$_{\rm TDB}$ 2460067.6617 + 27.36152\;$\times$\;$E$, with $E$ being the epoch number relative to epoch 53. This simple model does not fit epochs 54 and 57 well, and the main periodicity of the TTVs is probably longer than the 3.7~year-long sine wave period. There are tentative indications of "chopping" \citep{deck2015}, for example the "jump" between transit epochs 53 and 54, but more transit observations are clearly needed to characterize the TTVs.

\begin{table}
  \setlength{\tabcolsep}{5pt}
\renewcommand{\arraystretch}{1.25}
\centering
\caption{Sine wave modeling of the TTVs.}\label{table.TTVsSine}
\begin{tabular}{lcc}
\hline
\hline
Parameter & Units &  Median and 68.3\% CI  \\
\hline
$T_0$                  & [BJD$_{\mathrm{TDB}}$] & $2460067.6622\pm0.0027$ \\
Orbital period, $P$    & [days]                    & $27.36155\pm0.00015$ \\
Sine $t_s$             & [epochs]                 & $-15.62\pm0.38$ \\
Sine period, $P_s$           & [epochs]                 & $49.3\pm2.7$ \\
Sine amplitude, $A$         & [minutes]              & $40.1\pm2.8$\\
\hline
\end{tabular}
\tablefoot{The model transit time is $T_0 + E\,P + A \sin{\left[\frac{2\pi (E - t_s)}{P_s} + \pi\right]}$, where $E$ is the epoch number relative to epoch 53.}
\end{table}

\subsection{Nonparametric model}

Although a sinusoidal model may be an adequate model, it risks lacking the flexibility to capture the intricacies produced by the actual planetary dynamics. On the other hand, N-body simulations are costly and often quite slow. In any case, a model correctly accounting for TTVs is needed to correctly estimate the planetary ephemeris.

For this subsection, we used a nonparametric model for the TTVs. More concretely, the variations with respect to a constant linear ephemeris were modeled using GP regression \citep{rasmussenwilliams2005, rajpaul2015}. The covariance of the model is given by a kernel function, whose hyperparameters were optimized simultaneously to the determination of the ephemeris parameters, $P$ and $T_0$.

Several kernel functions were tried. Leveraging the fact that the sum and product of two valid kernel functions is also a valid kernel function, we constructed a series of models based on the following: 

\begin{enumerate}
\item{the squared exponential function -- the radial basis function (RBF),

\begin{equation}
\label{eq.rbf}
k(t_i, t_j) = A^2 \exp\left(- \frac{d(t_i, t_j)^2}{2\tau^2} \right)\;\; ,
\end{equation}
where $A^2$ is the covariance amplitude, the variable $t_i$ is the epoch of $i$-th transit epoch, and $d$ is a function computing the distance, here $d(t_i, t_j) = t_i - t_j$;}

\item{the exp-sine function (i.e., a strictly periodic kernel),

\begin{equation}
\label{eq.expsine}
k(t_i, t_j) = A^2 \exp\left(-
        \frac{ 2\sin^2(\pi\,d(t_i, t_j)/\mathcal{P}) }{ \epsilon^ 2} \right)\;\;,
\end{equation}
        
where $\mathcal{P}$ is the covariance periodicity; and} 

\item{a quasi-periodic function (i.e., the product of the squared exponential and the exp-sine kernels).}
\end{enumerate}
        
In addition, we explored the inclusion of a white noise term in the covariance (i.e., a diagonal kernel function). Readers can refer to \citet{rasmussenwilliams2005} for a description of each kernel function.

Vague priors were set for the parameters of the linear mean model, $P$ and $T_0$, by using a normal prior with large variance. In this manner, the marginal likelihood of the model can be computed analytically. The marginal likelihood thus computed was optimized with respect to the hyperparameters using the limited-memory Broyden–Fletcher–Goldfarb–Shanno (LM-BFGS) algorithm implemented in the Python \scipy package \optimize \citep{scipy}.

Overall, we found that kernels without a periodic term (the RBF kernel) produced worse results than those with a periodic term. Also, quasi-periodic kernels have very long decay times, in agreement with the fact that -- at most -- a single period is covered by the current observations. In fact, the periodic and quasi-periodic solutions were identical in all aspects, and are therefore not reported. 

If the kernel functions enumerated above are combined with a long-term trend modeled by summing an additional RBF function,
\begin{equation}
\label{eq.long}
A_\text{long} \exp \left(- \frac{d(t_i, t_j)^2}{\tau_\text{long}^2}\right)\;\;,
    \end{equation}
the values of the marginal likelihood are maximized, but the error in the mean orbital period is increased significantly. These solutions, though favored by the data, imply that the currently observed transits are only covering a small fraction of the much longer evolution with timescale $\tau_\text{long}$ of about 185 transit epochs, that is, over 13 years.

The results of all the tested kernel functions are listed in order of increasing optimized marginal likelihood in Table~\ref{table.gpttvs}. The value of the selected parameters are presented. We remind the reader that the values of the hyperparameters are optimized, and that they should therefore be considered with caution. On the other hand, for the period and the epoch of the transits of TOI-1227\,b, we have full Gaussian posterior distributions, whose summaries are provided. However, these distributions are conditional on the values of the hyperparameters.

\begin{figure}
\includegraphics[width=\textwidth/2]{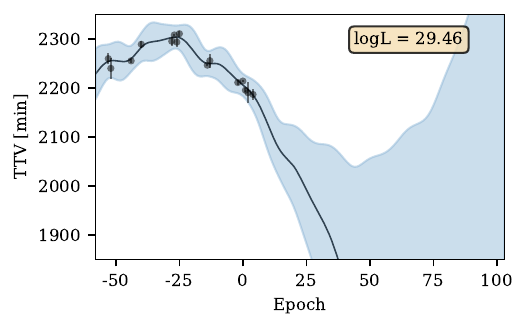}
\includegraphics[width=\textwidth/2]{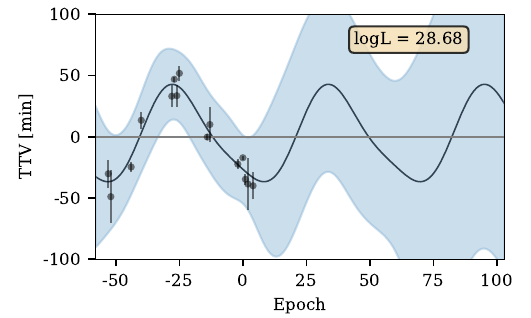}
\includegraphics[width=\textwidth/2]{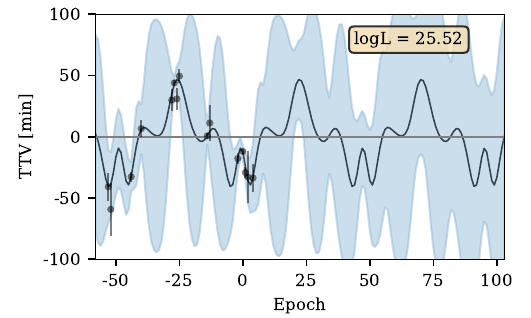}
\caption{ GP regression of the time variations in excess of mean linear ephemeris. The results are shown for kernels "per + long + white" (top), "per + white" (middle), and "per" (bottom). The black curve represents the posterior GP mean and the blue-shaded area represents the 3-$\sigma$ region for each epoch. The epoch number is relative to epoch 53. \label{fig.gpttvs}}
\end{figure}

We note that the solutions from kernels including a white noise component (+ white) imply an additional white noise term with an amplitude between 6.4 and 8.3 minutes. This may seem excessive in view of the current data, but our objective for this work was to avoid constraining the model drastically by setting bounds on the optimization. We also present the alternative models without a diagonal term added to the model covariance. In all cases, the uncertainty of the period and epoch of transits increased significantly with respect to the values in Table~\ref{table.TTVsSine}, because this approach adequately\footnote{This ignores the fact that the hyperparameter values are optimized. A full model would provide posterior distributions for these parameters as well. The approach here is called type-II maximum likelihood estimation.} propagates the uncertainty of the TTV model into the ephemeris. In Fig.~\ref{fig.gpttvs} we present the best model (per + long + white), the best model without a long-term trend (per + white), and the best model with neither a long-term trend nor white noise component (per).

\begin{table*}
\caption{Results of the GP regression model to the transit times.\label{table.gpttvs}}
\begin{tabular}{lrrrrrrrr}
 \hline\hline
 kernel\tablefootmark{a} & period, $P$ & $\sigma_P$ & $T_0$ & $\sigma_{T_0}$ & cov.~amplitude, $A$ & cov.~periodicity, $\mathcal{P}$ & $\sigma_\text{white}$ & logL \\
 & [d] & [d] & \multicolumn{2}{r}{[BJD$_\text{TDB}$ - 2,460,000]} & [min] & [epochs] & [min]\\
 \hline
RBF & 27.36090 & 0.00070 & 67.636 & 0.023 & 41.63 & -- & -- & 23.25 \\
per & 27.36109 & 0.00014 & 67.648 & 0.010 & 29.85 & 48.02 & -- & 25.52 \\
RBF + white & 27.36108 & 0.00086 & 67.637 & 0.030 & 48.69 & -- & 8.3 & 28.55 \\
per + white & 27.36130 & 0.00025 & 67.651 & 0.043 & 67.77 & 61.47 & 8.0 & 28.68 \\
per + long & 27.36172 & 0.01256 & 66.218 & 1.968 & 12.38 & 29.92 & -- & 29.01 \\
per + long + white & 27.36207 & 0.01324 & 66.102 & 2.131 & 15.40 & 29.56 & 6.4 & 29.46 \\
\hline
\end{tabular}
\tablefoottext{a}{Here "RBF" refer to the radial basis function (squared exponential; Eq.~\ref{eq.rbf}); "per" to the exp-sine function (Eq.~\ref{eq.expsine}); "long" to the additional long-term (i.e., long decay time) function (Eq.~\ref{eq.long}; "white" refers to an additional term to the covariance diagonal elements.}
\end{table*}

\subsection{Dynamical modeling}\label{section.nbody}

We have attempted to develop a physical model of the TTVs, exploring inner and outer perturbers around the first and second order mean motion resonances (exploring period ratios from 3:1 to 6:5). We assumed coplanar orbits and neglected the light-time effect \citep{irwin1952}, which is very small at these periods. We used the \rebound n-body code \citep{rein2012} with the \whf integrator \citep{rein2015} and an integration step of 0.01~days. We did not impose a stability criterion. The model was parameterized using, for each planet, the planet-to-star mass ratio, a pseudo-period\footnote{$P' \equiv \sqrt{\frac{4\pi^2a^{3}}{G M_{\star}}}$, with $a$ being the semi-major axis, $\mathcal G$ the gravitational constant, and $M_{\star}$ the mass of the star.}, the time of a reference conjunction (star, planet, and observer), and two products of the square root of the eccentricity: one multiplied by the cosine of the argument of periastron, and another multiplied by the sine of the argument of periastron. We used uniform priors for all parameters. The joint posterior distribution was sampled using the \emcee algorithm. We started the walkers for a 10~\MEarth mass for TOI-1227\,b, a plausible value following \citet{mann2022}, and a perturber offset from the period commensurability in accordance with a TTVs' super-period \citep{lithwick2012} of 3.7~years. 

\begin{figure*}
  \centering
  \includegraphics[width=0.95\textwidth]{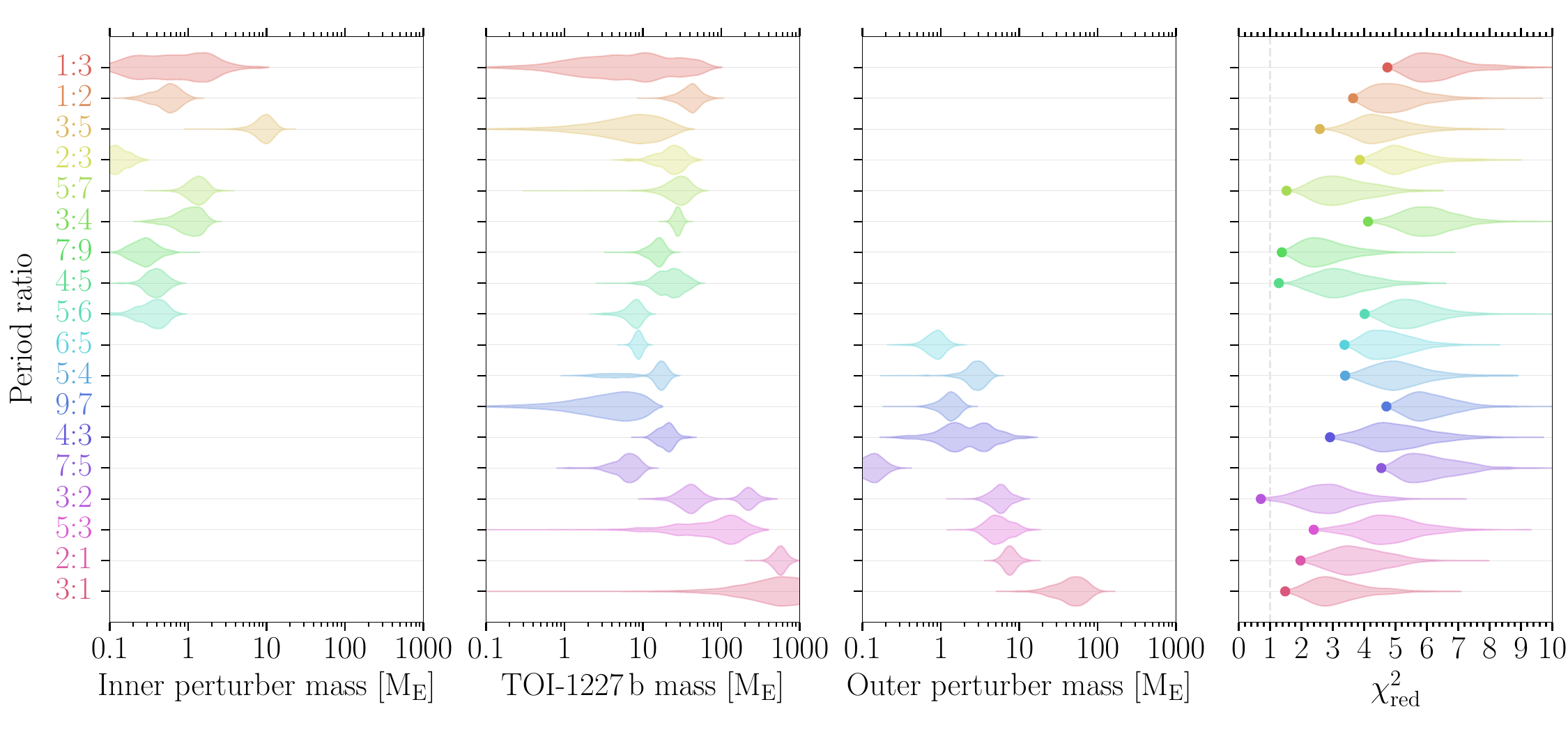}
  \caption{Exploration of perturbers near the inner and outer first and second order resonances (the period ratio is denoted in the format "perturber:TOI-1227\,b"). From left to right: Violin plots representing the marginal posterior of the mass for the inner perturber, the mass of TOI-1227\,b, the mass of the outer perturber, and the goodness of fit ($\chi^2_{\rm red}$), with the MAP value displayed as a dot.} \label{fig.MMR}
\end{figure*}

While we have not explored the entire parameter space of these orbital configurations, we have already found multiple very distinct solutions that provide a satisfactory fit. Figure~\ref{fig.MMR} shows the posterior for the perturber and TOI-1227\,b masses and a metric for the goodness of fit ($\chi^2_{\rm red}$). 
The best fit was obtained for an exterior 3:2 MMR. In this case, the TOI-1227 b mass distribution is bimodal, with a peak at $40\ \mathrm{M}_\oplus$ and a perturber of mass $5.7\pm1.7\ \mathrm{M}_\oplus$. The next best fits correspond to the inner resonances 5:7, 7:9, and 4:5. However, these fits show an indication of overfitting, with very small masses for the perturber and very constrained mass ranges for TOI-1227 b. These configurations could be ruled out with a longer TTV coverage, resolving the super-period.

\begin{figure}
  \centering
  \includegraphics[width=0.49\textwidth]{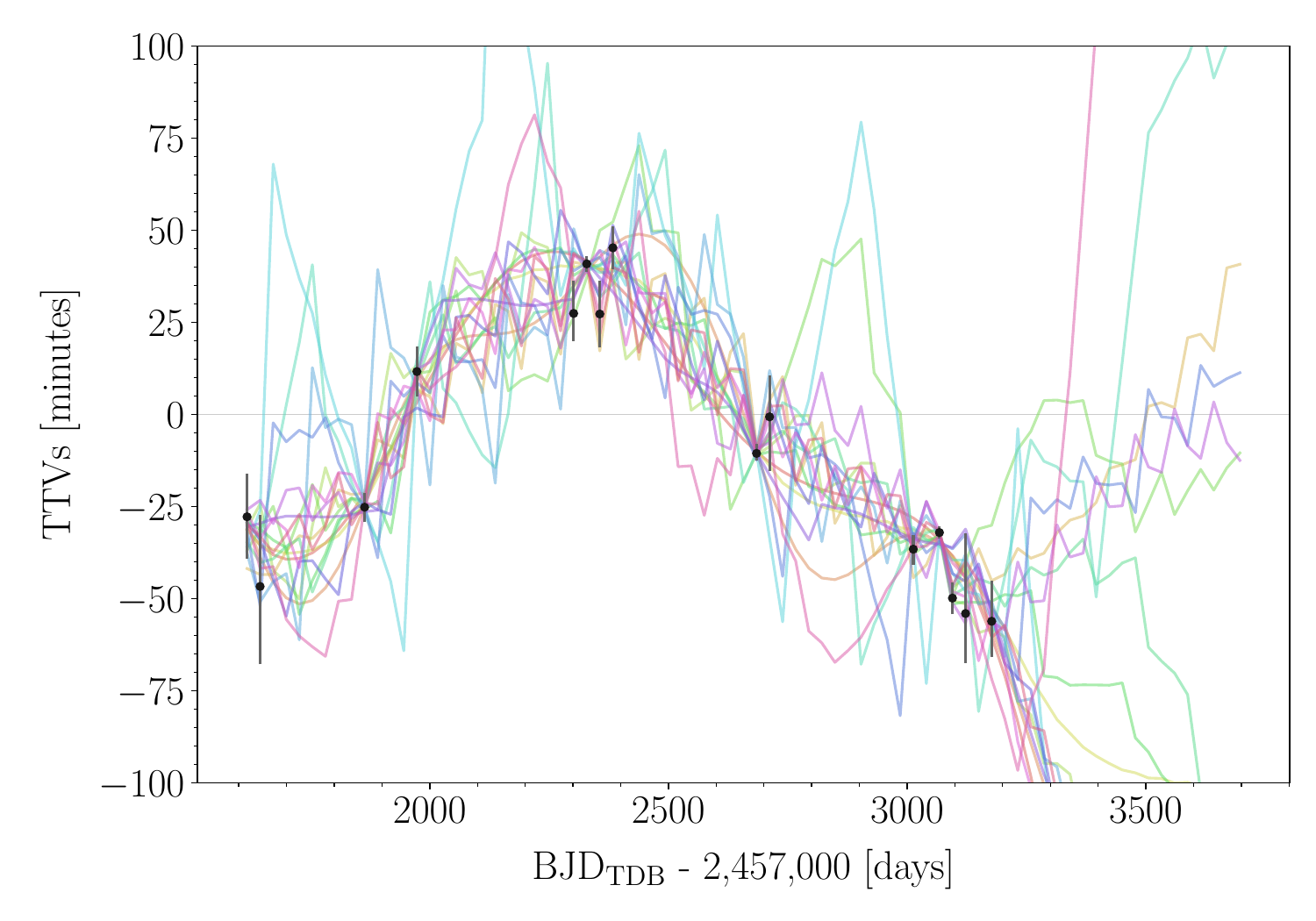}
  \caption{Same as the upper panel of Fig.~\ref{fig.TTVs} but showing the MAP models of the exploration of the perturbers close to the inner and outer first and second order resonances. The color code is the same as in Fig.~\ref{fig.MMR}.} \label{fig.TTVsMMR}
\end{figure}

The MAP model for each period ratio tested is plotted in Fig.~\ref{fig.TTVsMMR}. Even with the simplistic assumption of coplanar orbits, the dynamical model is flexible enough to capture the observed TTVs, including the jump between epochs 53 and 54. This exercise proves that the TTVs can be reasonably fitted with several orbital configurations, and thus that the problem is degenerate with the current dataset. We again conclude that more data are needed to find the right solution. The problem naturally becomes more complex if additional dynamically relevant perturbers are present in the system.

\section{Results and discussion}\label{section.results}

We have detected significant TTVs of the 11~Myr exoplanet TOI-1227\,b. Those TTVs are well above what could be explained by spots (TOI-1227 shows rotational modulation in the TESS light curve with a period of 1.65~days): they amount to 14\% of the transit duration (see Fig.~\ref{fig.model}), while spots can cause TTVs of $\sim$1\% of the transit duration at most \citep{ioannidis2016}. We conclude that the observed TTVs are most likely caused by at least one additional planet in the system, but additional transit observations are needed to constrain the orbital configuration.

One probable perturber could be the external 3:2, with a mass of $5.7\pm1.7$~\MEarth. Among the configurations tested, the external 3:2 perturber provided the best fit to the data. Interestingly, capture into the 3:2 is the typical expected outcome of convergent type~I migration for sub-Neptunes \citep{Kajtazi2023}. In addition, the configuration is stable during the migration phase as the outer perturber is less massive than the inner one \citep{Deck2015a}. Unfortunately, if this perturber is coplanar with TOI-1227\,b, it will not transit.

In addition to detecting TTVs, the additional transit observations slightly improve the transit parameters derived in \citet{mann2022}. Specifically, grazing transits are now ruled out, eliminating the tail of high planet-to-star radius ratios in the posterior. The planetary radius we derived in ($0.854\pm0.049$~\Rjup) precisely matches the one of \citet{mann2022}, $0.854^{+0.067}_{-0.052}$~\Rjup, with just slightly improved error bars. While the planet-to-star radius ratio now has a 1.9\% uncertainty, the planetary radius is uncertain by 5.7\%, dominated by the 5.4\% uncertainty on the stellar radius. The planet-to-star radius ratio measurement may also be biased, as it assumes an unspotted star \citep{czesla2009}.

We expanded the transit chromatic analysis of \citet{mann2022}, and again found no significant chromaticity. It is anticipated that depths will increase at optical wavelengths due to the presence of unocculted spots. However, the g' measurement suggests that spots may not significantly influence the measurement of the planet-to-star radius ratio. An alternative interpretation could be that the spots are distributed semi-uniformly. This implies that there is roughly an equal number of unocculted and occulted spots and, as a result, their effects would largely neutralize each other.

Based on our understanding, there are no false positive scenarios that could account for the observed 40-minute TTVs of TOI-1227\,b. As a result, these observations appear to unequivocally confirm TOI-1227\,b as a planet.
We encourage intensive observations of TOI-1227\,b's transits, as they could lead to a measurement of the planet's mass.

TOI-1227\,b provides a rare opportunity for transmission spectroscopy of a still-contracting young exoplanet using JWST \citep{gardner2006} or high-resolution spectrographs, although starspots may pose a challenge \citep{rackham2018}. The transmission spectroscopy metric \citep[TSM,][]{kempton2018} could range from 95 to 950 for a mass of 50 and 5~\MEarth, respectively. However, before attempting this observation, we should improve the forecast of TOI-1227\,b transits. Additionally, measuring the mass of TOI-1227 b would aid in better interpreting the transmission spectrum.

\begin{acknowledgements}

We are grateful to the ESO/La Silla staff for their continuous support of ExTrA and to the wintering staff at Concordia Station in Antarctica who look after ASTEP during the long austral winter night. We acknowledge funding from the European Research Council under the ERC Grant Agreement n. 337591-ExTrA. 

This paper includes data collected with the TESS mission, obtained from the MAST data archive at the Space Telescope Science Institute (STScI). Funding for the TESS mission is provided by the NASA Explorer Program. STScI is operated by the Association of Universities for Research in Astronomy, Inc., under NASA contract NAS 5–26555. We acknowledge the use of public TESS data from the pipelines at the TESS Science Office and at the TESS Science Processing Operations Center. Resources supporting this work were provided by the NASA High-End Computing (HEC) program through the NASA Advanced Supercomputing (NAS) Division at Ames Research Center for the production of the SPOC data products. 

This work makes use of observations from the Las Cumbres Observatory global telescope network.

Simulations in this paper made use of the \rebound code which can be downloaded freely at \texttt{http://github.com/hannorein/rebound}. 

We thank the Swiss National Science Foundation (SNSF) and the Geneva University for their continuous support to our planet search programs. This work has been carried out within the framework of the National Centre of Competence in Research PlanetS supported by the Swiss National Science Foundation under grants 51NF40\_182901 and 51NF40\_205606. The authors acknowledge the financial support of the SNSF.

This paper is based on data collected by the SPECULOOS-South Observatory at the ESO Paranal Observatory in Chile. The ULiege's contribution to SPECULOOS has received funding from the European Research Council under the European Union's Seventh Framework Programme (FP/2007-2013) (grant Agreement n$^\circ$ 336480/SPECULOOS), from the Balzan Prize and Francqui Foundations, from the Belgian Scientific Research Foundation (F.R.S.-FNRS; grant n$^\circ$ T.0109.20), from the University of Li\`ege, and from the ARC grant for Concerted Research Actions financed by the Wallonia-Brussels Federation. The Cambridge contribution is supported by a grant from the Simons Foundation (PI Queloz, grant number 327127).
The Birmingham contribution research is in part funded by the European Union's Horizon 2020 research and innovation programme (grant's agreement n$^{\circ}$ 803193/BEBOP), from the MERAC foundation, and from the Science and Technology Facilities Council (STFC; grant n$^\circ$ ST/S00193X/1, and ST/W000385/1).
This is also based on data collected by the TRAPPIST-South telescope at the ESO La Silla Observatory. TRAPPIST is funded by the Belgian Fund for Scientific Research (Fond National de la Recherche Scientifique, FNRS) under the grant FRFC 2.5.594.09.F, with the participation of the Swiss National Science Fundation (SNF). EJ is a Senior Research Scientist at FNRS.

This paper uses data obtained with the ASTEP telescope, at Concordia Station in Antarctica. ASTEP benefited from the support of the
French and Italian polar agencies IPEV and PNRA in
the framework of the Concordia station program, from
OCA, INSU, Idex UCAJEDI (ANR- 15-IDEX-01) and
ESA through the Science Faculty of the European Space Research and Technology Centre (ESTEC).
The Birmingham contribution to ASTEP is supported by the European Union's Horizon 2020 research and innovation programme (grant's agreement n$^{\circ}$ 803193/BEBOP), and from the Science and Technology Facilities Council (STFC; grant n$^\circ$ ST/S00193X/1, and ST/W002582/1). 

This research has made use of the Spanish Virtual Observatory (https://svo.cab.inta-csic.es) project funded by MCIN/AEI/10.13039/501100011033/ through grant PID2020-112949GB-I00.

This work was supported by the "Programme National de Physique Stellaire" (PNPS) and "Programme National de Plan\'etology" of CNRS/INSU co-funded by CEA and CNES

M.L. acknowledges support of the Swiss National Science Foundation under grant number PCEFP2\_194576.

J.V. acknowledges support from the Swiss National Science Foundation (SNSF) under grant PZ00P2\_208945.

\end{acknowledgements}

\bibliographystyle{aa}
\bibliography{TOI-1227}

\begin{thebibliography}{80}
\expandafter\ifx\csname natexlab\endcsname\relax\def\natexlab#1{#1}\fi

\bibitem[{{Abe} {et~al.}(2013){Abe}, {Gon{\c{c}}alves}, {Agabi}, {Alapini},
  {Guillot}, {M{\'e}karnia}, {Rivet}, {Schmider}, {Crouzet}, {Fortney}, {Pont},
  {Barbieri}, {Daban}, {Fante{\"\i}-Caujolle}, {Gouvret}, {Bresson}, {Roussel},
  {Bonhomme}, {Robini}, {Dugu{\'e}}, {Bondoux}, {P{\'e}ron}, {Petit},
  {Szul{\'a}gyi}, {Fruth}, {Erikson}, {Rauer}, {Fressin}, {Valbousquet},
  {Blanc}, {Le van Suu}, \& {Aigrain}}]{Abe2013}
{Abe}, L., {Gon{\c{c}}alves}, I., {Agabi}, A., {et~al.} 2013, \aap, 553, A49

\bibitem[{{Agol} {et~al.}(2005){Agol}, {Steffen}, {Sari}, \&
  {Clarkson}}]{agol2005}
{Agol}, E., {Steffen}, J., {Sari}, R., \& {Clarkson}, W. 2005, \mnras, 359, 567

\bibitem[{{Bolmont} {et~al.}(2017){Bolmont}, {Gallet}, {Mathis}, {Charbonnel},
  {Amard}, \& {Alibert}}]{Bolmont17}
{Bolmont}, E., {Gallet}, F., {Mathis}, S., {et~al.} 2017, \aap, 604, A113

\bibitem[{{Bonfils} {et~al.}(2015){Bonfils}, {Almenara}, {Jocou}, {Wunsche},
  {Kern}, {Delboulb{\'e}}, {Delfosse}, {Feautrier}, {Forveille}, {Gluck},
  {Lafrasse}, {Magnard}, {Maurel}, {Moulin}, {Murgas}, {Rabou}, {Rochat},
  {Roux}, \& {Stadler}}]{Bon2015}
{Bonfils}, X., {Almenara}, J.~M., {Jocou}, L., {et~al.} 2015, in Society of
  Photo-Optical Instrumentation Engineers (SPIE) Conference Series, Vol. 9605,
  Techniques and Instrumentation for Detection of Exoplanets VII, ed.
  S.~{Shaklan}, 96051L

\bibitem[{{Caldwell} {et~al.}(2020){Caldwell}, {Tenenbaum}, {Twicken},
  {Jenkins}, {Ting}, {Smith}, {Hedges}, {Fausnaugh}, {Rose}, \&
  {Burke}}]{caldwell2020}
{Caldwell}, D.~A., {Tenenbaum}, P., {Twicken}, J.~D., {et~al.} 2020, Research
  Notes of the American Astronomical Society, 4, 201

\bibitem[{{Cale} {et~al.}(2021){Cale}, {Reefe}, {Plavchan}, {Tanner}, {Gaidos},
  {Gagn{\'e}}, {Gao}, {Kane}, {B{\'e}jar}, {Lodieu}, {Anglada-Escud{\'e}},
  {Ribas}, {Pall{\'e}}, {Quirrenbach}, {Amado}, {Reiners}, {Caballero}, {Rosa
  Zapatero Osorio}, {Dreizler}, {Howard}, {Fulton}, {Xuesong Wang}, {Collins},
  {El Mufti}, {Wittrock}, {Gilbert}, {Barclay}, {Klein}, {Martioli},
  {Wittenmyer}, {Wright}, {Addison}, {Hirano}, {Tamura}, {Kotani}, {Narita},
  {Vermilion}, {Lee}, {Geneser}, {Teske}, {Quinn}, {Latham}, {Esquerdo},
  {Calkins}, {Berlind}, {Zohrabi}, {Stibbards}, {Kotnana}, {Jenkins},
  {Twicken}, {Henze}, {Kidwell}, {Burke}, {Villase{\~n}or}, \&
  {Boyd}}]{cale2021}
{Cale}, B.~L., {Reefe}, M., {Plavchan}, P., {et~al.} 2021, \aj, 162, 295

\bibitem[{{Czesla} {et~al.}(2009){Czesla}, {Huber}, {Wolter}, {Schr{\"o}ter},
  \& {Schmitt}}]{czesla2009}
{Czesla}, S., {Huber}, K.~F., {Wolter}, U., {Schr{\"o}ter}, S., \& {Schmitt},
  J.~H.~M.~M. 2009, \aap, 505, 1277

\bibitem[{{David} {et~al.}(2016){David}, {Hillenbrand}, {Petigura},
  {Carpenter}, {Crossfield}, {Hinkley}, {Ciardi}, {Howard}, {Isaacson}, {Cody},
  {Schlieder}, {Beichman}, \& {Barenfeld}}]{david2016}
{David}, T.~J., {Hillenbrand}, L.~A., {Petigura}, E.~A., {et~al.} 2016, \nat,
  534, 658

\bibitem[{{David} {et~al.}(2019){David}, {Petigura}, {Luger}, {Foreman-Mackey},
  {Livingston}, {Mamajek}, \& {Hillenbrand}}]{david2019}
{David}, T.~J., {Petigura}, E.~A., {Luger}, R., {et~al.} 2019, \apjl, 885, L12

\bibitem[{{Deck} \& {Agol}(2015)}]{deck2015}
{Deck}, K.~M. \& {Agol}, E. 2015, \apj, 802, 116

\bibitem[{Deck \& Batygin(2015)}]{Deck2015a}
Deck, K.~M. \& Batygin, K. 2015, The Astrophysical Journal, 810, 119

\bibitem[{Delisle \& Laskar(2014)}]{Delisle2014}
Delisle, J.-B. \& Laskar, J. 2014, Astronomy and Astrophysics, 570, L7

\bibitem[{{Delrez} {et~al.}(2018){Delrez}, {Gillon}, {Queloz}, {Demory},
  {Almleaky}, {de Wit}, {Jehin}, {Triaud}, {Barkaoui}, {Burdanov}, {Burgasser},
  {Ducrot}, {McCormac}, {Murray}, {Silva Fernandes}, {Sohy}, {Thompson}, {Van
  Grootel}, {Alonso}, {Benkhaldoun}, \& {Rebolo}}]{Delrez2018}
{Delrez}, L., {Gillon}, M., {Queloz}, D., {et~al.} 2018, in Society of
  Photo-Optical Instrumentation Engineers (SPIE) Conference Series, Vol. 10700,
  Ground-based and Airborne Telescopes VII, ed. H.~K. {Marshall} \&
  J.~{Spyromilio}, 107001I

\bibitem[{{Donati} {et~al.}(2023){Donati}, {Cristofari}, {Finociety}, {Klein},
  {Moutou}, {Gaidos}, {Cadieux}, {Artigau}, {Correia}, {Bou{\'e}}, {Cook},
  {Carmona}, {Lehmann}, {Bouvier}, {Martioli}, {Morin}, {Fouqu{\'e}},
  {Delfosse}, {Doyon}, {H{\'e}brard}, {Alencar}, {Laskar}, {Arnold}, {Petit},
  {K{\'o}sp{\'a}l}, {Vidotto}, {Folsom}, \& {collaboration}}]{donati2023}
{Donati}, J.~F., {Cristofari}, P.~I., {Finociety}, B., {et~al.} 2023, \mnras,
  525, 455

\bibitem[{{Dorn} {et~al.}(2017){Dorn}, {Venturini}, {Khan}, {Heng}, {Alibert},
  {Helled}, {Rivoldini}, \& {Benz}}]{Dorn_2017}
{Dorn}, C., {Venturini}, J., {Khan}, A., {et~al.} 2017, \aap, 597, A37

\bibitem[{{Dransfield} {et~al.}(2022){Dransfield}, {M{\'e}karnia}, {Triaud},
  {Guillot}, {Abe}, {Garc{\'\i}a}, {Timmermans}, {Crouzet}, {Schmider},
  {Agabi}, {Suarez}, {Bendjoya}, {Guenther}, {Lai}, {Merin}, \&
  {Stee}}]{Dransfield2022}
{Dransfield}, G., {M{\'e}karnia}, D., {Triaud}, A. H.~M.~J., {et~al.} 2022, in
  Society of Photo-Optical Instrumentation Engineers (SPIE) Conference Series,
  Vol. 12186, Observatory Operations: Strategies, Processes, and Systems IX,
  ed. D.~S. {Adler}, R.~L. {Seaman}, \& C.~R. {Benn}, 121861F

\bibitem[{{Espinoza}(2018)}]{espinoza2018}
{Espinoza}, N. 2018, Research Notes of the American Astronomical Society, 2,
  209

\bibitem[{{Espinoza} {et~al.}(2019){Espinoza}, {Kossakowski}, \&
  {Brahm}}]{espinoza2019}
{Espinoza}, N., {Kossakowski}, D., \& {Brahm}, R. 2019, \mnras, 490, 2262

\bibitem[{{Finociety} {et~al.}(2023){Finociety}, {Donati}, {Cristofari},
  {Moutou}, {Cadieux}, {Cook}, {Artigau}, {Baruteau}, {Debras}, {Fouqu{\'e}},
  {Bouvier}, {Alencar}, {Delfosse}, {Grankin}, {Carmona}, {Petit},
  {K{\'o}sp{\'a}l}, \& {The Sls/Spice Consortium}}]{finociety2023}
{Finociety}, B., {Donati}, J.~F., {Cristofari}, P.~I., {et~al.} 2023, \mnras,
  526, 4627

\bibitem[{{Foreman-Mackey} {et~al.}(2017){Foreman-Mackey}, {Agol},
  {Ambikasaran}, \& {Angus}}]{foreman-mackey2017}
{Foreman-Mackey}, D., {Agol}, E., {Ambikasaran}, S., \& {Angus}, R. 2017, \aj,
  154, 220

\bibitem[{{Foreman-Mackey} {et~al.}(2013){Foreman-Mackey}, {Hogg}, {Lang}, \&
  {Goodman}}]{emcee}
{Foreman-Mackey}, D., {Hogg}, D.~W., {Lang}, D., \& {Goodman}, J. 2013, \pasp,
  125, 306

\bibitem[{{Gaia Collaboration} {et~al.}(2018){Gaia Collaboration}, {Brown},
  {Vallenari}, {Prusti}, {de Bruijne}, {Babusiaux}, {Bailer-Jones}, {Biermann},
  {Evans}, {Eyer}, {Jansen}, {Jordi}, {Klioner}, {Lammers}, {Lindegren},
  {Luri}, {Mignard}, {Panem}, {Pourbaix}, {Randich}, {Sartoretti}, {Siddiqui},
  {Soubiran}, {van Leeuwen}, {Walton}, {Arenou}, {Bastian}, {Cropper},
  {Drimmel}, {Katz}, {Lattanzi}, {Bakker}, {Cacciari}, {Casta{\~n}eda},
  {Chaoul}, {Cheek}, {De Angeli}, {Fabricius}, {Guerra}, {Holl}, {Masana},
  {Messineo}, {Mowlavi}, {Nienartowicz}, {Panuzzo}, {Portell}, {Riello},
  {Seabroke}, {Tanga}, {Th{\'e}venin}, {Gracia-Abril}, {Comoretto},
  {Garcia-Reinaldos}, {Teyssier}, {Altmann}, {Andrae}, {Audard},
  {Bellas-Velidis}, {Benson}, {Berthier}, {Blomme}, {Burgess}, {Busso},
  {Carry}, {Cellino}, {Clementini}, {Clotet}, {Creevey}, {Davidson}, {De
  Ridder}, {Delchambre}, {Dell'Oro}, {Ducourant},
  {Fern{\'a}ndez-Hern{\'a}ndez}, {Fouesneau}, {Fr{\'e}mat}, {Galluccio},
  {Garc{\'\i}a-Torres}, {Gonz{\'a}lez-N{\'u}{\~n}ez}, {Gonz{\'a}lez-Vidal},
  {Gosset}, {Guy}, {Halbwachs}, {Hambly}, {Harrison}, {Hern{\'a}ndez},
  {Hestroffer}, {Hodgkin}, {Hutton}, {Jasniewicz}, {Jean-Antoine-Piccolo},
  {Jordan}, {Korn}, {Krone-Martins}, {Lanzafame}, {Lebzelter}, {L{\"o}ffler},
  {Manteiga}, {Marrese}, {Mart{\'\i}n-Fleitas}, {Moitinho}, {Mora}, {Muinonen},
  {Osinde}, {Pancino}, {Pauwels}, {Petit}, {Recio-Blanco}, {Richards},
  {Rimoldini}, {Robin}, {Sarro}, {Siopis}, {Smith}, {Sozzetti}, {S{\"u}veges},
  {Torra}, {van Reeven}, {Abbas}, {Abreu Aramburu}, {Accart}, {Aerts},
  {Altavilla}, {{\'A}lvarez}, {Alvarez}, {Alves}, {Anderson}, {Andrei},
  {Anglada Varela}, {Antiche}, {Antoja}, {Arcay}, {Astraatmadja}, {Bach},
  {Baker}, {Balaguer-N{\'u}{\~n}ez}, {Balm}, {Barache}, {Barata}, {Barbato},
  {Barblan}, {Barklem}, {Barrado}, {Barros}, {Barstow}, {Bartholom{\'e}
  Mu{\~n}oz}, {Bassilana}, {Becciani}, {Bellazzini}, {Berihuete}, {Bertone},
  {Bianchi}, {Bienaym{\'e}}, {Blanco-Cuaresma}, {Boch}, {Boeche}, {Bombrun},
  {Borrachero}, {Bossini}, {Bouquillon}, {Bourda}, {Bragaglia}, {Bramante},
  {Breddels}, {Bressan}, {Brouillet}, {Br{\"u}semeister}, {Brugaletta},
  {Bucciarelli}, {Burlacu}, {Busonero}, {Butkevich}, {Buzzi}, {Caffau},
  {Cancelliere}, {Cannizzaro}, {Cantat-Gaudin}, {Carballo}, {Carlucci},
  {Carrasco}, {Casamiquela}, {Castellani}, {Castro-Ginard}, {Charlot},
  {Chemin}, {Chiavassa}, {Cocozza}, {Costigan}, {Cowell}, {Crifo}, {Crosta},
  {Crowley}, {Cuypers}, {Dafonte}, {Damerdji}, {Dapergolas}, {David}, {David},
  {de Laverny}, {De Luise}, {De March}, {de Martino}, {de Souza}, {de Torres},
  {Debosscher}, {del Pozo}, {Delbo}, {Delgado}, {Delgado}, {Di Matteo},
  {Diakite}, {Diener}, {Distefano}, {Dolding}, {Drazinos}, {Dur{\'a}n},
  {Edvardsson}, {Enke}, {Eriksson}, {Esquej}, {Eynard Bontemps}, {Fabre},
  {Fabrizio}, {Faigler}, {Falc{\~a}o}, {Farr{\`a}s Casas}, {Federici},
  {Fedorets}, {Fernique}, {Figueras}, {Filippi}, {Findeisen}, {Fonti},
  {Fraile}, {Fraser}, {Fr{\'e}zouls}, {Gai}, {Galleti}, {Garabato},
  {Garc{\'\i}a-Sedano}, {Garofalo}, {Garralda}, {Gavel}, {Gavras}, {Gerssen},
  {Geyer}, {Giacobbe}, {Gilmore}, {Girona}, {Giuffrida}, {Glass}, {Gomes},
  {Granvik}, {Gueguen}, {Guerrier}, {Guiraud}, {Guti{\'e}rrez-S{\'a}nchez},
  {Haigron}, {Hatzidimitriou}, {Hauser}, {Haywood}, {Heiter}, {Helmi}, {Heu},
  {Hilger}, {Hobbs}, {Hofmann}, {Holland}, {Huckle}, {Hypki}, {Icardi},
  {Jan{\ss}en}, {Jevardat de Fombelle}, {Jonker}, {Juh{\'a}sz}, {Julbe},
  {Karampelas}, {Kewley}, {Klar}, {Kochoska}, {Kohley}, {Kolenberg},
  {Kontizas}, {Kontizas}, {Koposov}, {Kordopatis}, {Kostrzewa-Rutkowska},
  {Koubsky}, {Lambert}, {Lanza}, {Lasne}, {Lavigne}, {Le Fustec}, {Le
  Poncin-Lafitte}, {Lebreton}, {Leccia}, {Leclerc}, {Lecoeur-Taibi},
  {Lenhardt}, {Leroux}, {Liao}, {Licata}, {Lindstr{\o}m}, {Lister}, {Livanou},
  {Lobel}, {L{\'o}pez}, {Managau}, {Mann}, {Mantelet}, {Marchal}, {Marchant},
  {Marconi}, {Marinoni}, {Marschalk{\'o}}, {Marshall}, {Martino}, {Marton},
  {Mary}, {Massari}, {Matijevi{\v{c}}}, {Mazeh}, {McMillan}, {Messina},
  {Michalik}, {Millar}, {Molina}, {Molinaro}, {Moln{\'a}r}, {Montegriffo},
  {Mor}, {Morbidelli}, {Morel}, {Morris}, {Mulone}, {Muraveva}, {Musella},
  {Nelemans}, {Nicastro}, {Noval}, {O'Mullane}, {Ord{\'e}novic},
  {Ord{\'o}{\~n}ez-Blanco}, {Osborne}, {Pagani}, {Pagano}, {Pailler},
  {Palacin}, {Palaversa}, {Panahi}, {Pawlak}, {Piersimoni}, {Pineau}, {Plachy},
  {Plum}, {Poggio}, {Poujoulet}, {Pr{\v{s}}a}, {Pulone}, {Racero}, {Ragaini},
  {Rambaux}, {Ramos-Lerate}, {Regibo}, {Reyl{\'e}}, {Riclet}, {Ripepi}, {Riva},
  {Rivard}, {Rixon}, {Roegiers}, {Roelens}, {Romero-G{\'o}mez}, {Rowell},
  {Royer}, {Ruiz-Dern}, {Sadowski}, {Sagrist{\`a} Sell{\'e}s}, {Sahlmann},
  {Salgado}, {Salguero}, {Sanna}, {Santana-Ros}, {Sarasso}, {Savietto},
  {Schultheis}, {Sciacca}, {Segol}, {Segovia}, {S{\'e}gransan}, {Shih},
  {Siltala}, {Silva}, {Smart}, {Smith}, {Solano}, {Solitro}, {Sordo}, {Soria
  Nieto}, {Souchay}, {Spagna}, {Spoto}, {Stampa}, {Steele},
  {Steidelm{\"u}ller}, {Stephenson}, {Stoev}, {Suess}, {Surdej}, {Szabados},
  {Szegedi-Elek}, {Tapiador}, {Taris}, {Tauran}, {Taylor}, {Teixeira},
  {Terrett}, {Teyssand ier}, {Thuillot}, {Titarenko}, {Torra Clotet}, {Turon},
  {Ulla}, {Utrilla}, {Uzzi}, {Vaillant}, {Valentini}, {Valette}, {van Elteren},
  {Van Hemelryck}, {van Leeuwen}, {Vaschetto}, {Vecchiato}, {Veljanoski},
  {Viala}, {Vicente}, {Vogt}, {von Essen}, {Voss}, {Votruba}, {Voutsinas},
  {Walmsley}, {Weiler}, {Wertz}, {Wevers}, {Wyrzykowski}, {Yoldas},
  {{\v{Z}}erjal}, {Ziaeepour}, {Zorec}, {Zschocke}, {Zucker}, {Zurbach}, \&
  {Zwitter}}]{gaia2018}
{Gaia Collaboration}, {Brown}, A.~G.~A., {Vallenari}, A., {et~al.} 2018, \aap,
  616, A1

\bibitem[{{Garcia} {et~al.}(2022){Garcia}, {Timmermans}, {Pozuelos}, {Ducrot},
  {Gillon}, {Delrez}, {Wells}, \& {Jehin}}]{garcia2022}
{Garcia}, L.~J., {Timmermans}, M., {Pozuelos}, F.~J., {et~al.} 2022, \mnras,
  509, 4817

\bibitem[{{Gardner} {et~al.}(2006){Gardner}, {Mather}, {Clampin}, {Doyon},
  {Greenhouse}, {Hammel}, {Hutchings}, {Jakobsen}, {Lilly}, {Long}, {Lunine},
  {McCaughrean}, {Mountain}, {Nella}, {Rieke}, {Rieke}, {Rix}, {Smith},
  {Sonneborn}, {Stiavelli}, {Stockman}, {Windhorst}, \& {Wright}}]{gardner2006}
{Gardner}, J.~P., {Mather}, J.~C., {Clampin}, M., {et~al.} 2006, \ssr, 123, 485

\bibitem[{Gillon {et~al.}(2011)Gillon, Jehin, Magain, Chantry,
  Hutsem{\'{e}}kers, Manfroid, Queloz, \& Udry}]{Gillon2011}
Gillon, M., Jehin, E., Magain, P., {et~al.} 2011, {EPJ} Web of Conferences, 11,
  06002

\bibitem[{Goodman \& Weare(2010)}]{goodmanweare2010}
Goodman, J. \& Weare, J. 2010, Communications in applied mathematics and
  computational science, 5, 65

\bibitem[{{Guillot}(2010)}]{Guillot10}
{Guillot}, T. 2010, \aap, 520, A27

\bibitem[{{Guillot} {et~al.}(2015){Guillot}, {Abe}, {Agabi}, {Rivet}, {Daban},
  {M{\'e}karnia}, {Aristidi}, {Schmider}, {Crouzet}, {Gon{\c{c}}alves},
  {Gouvret}, {Ottogalli}, {Faradji}, {Blanc}, {Bondoux}, \&
  {Valbousquet}}]{Guillot2015}
{Guillot}, T., {Abe}, L., {Agabi}, A., {et~al.} 2015, Astronomische
  Nachrichten, 336, 638

\bibitem[{{Gupta} \& {Schlichting}(2019)}]{Gupta19}
{Gupta}, A. \& {Schlichting}, H.~E. 2019, \mnras, 487, 24

\bibitem[{{Haldemann} {et~al.}(2023){Haldemann}, {Dorn}, {Venturini},
  {Alibert}, \& {Benz}}]{Haldemann23}
{Haldemann}, J., {Dorn}, C., {Venturini}, J., {Alibert}, Y., \& {Benz}, W.
  2023, Astronomy \& Astrophysics, in press

\bibitem[{{Holman} \& {Murray}(2005)}]{holman2005}
{Holman}, M.~J. \& {Murray}, N.~W. 2005, Science, 307, 1288

\bibitem[{{Ioannidis} {et~al.}(2016){Ioannidis}, {Huber}, \&
  {Schmitt}}]{ioannidis2016}
{Ioannidis}, P., {Huber}, K.~F., \& {Schmitt}, J.~H.~M.~M. 2016, \aap, 585, A72

\bibitem[{{Irwin}(1952)}]{irwin1952}
{Irwin}, J.~B. 1952, \apj, 116, 211

\bibitem[{{Jehin} {et~al.}(2018){Jehin}, {Gillon}, {Queloz}, {Delrez},
  {Burdanov}, {Murray}, {Sohy}, {Ducrot}, {Sebastian}, {Thompson}, {McCormac},
  {Almleaky}, {Burgasser}, {Demory}, {de Wit}, {Barkaoui}, {Pozuelos},
  {Triaud}, \& {Grootel}}]{Jehin2018Msngr}
{Jehin}, E., {Gillon}, M., {Queloz}, D., {et~al.} 2018, The Messenger, 174, 2

\bibitem[{{Jehin} {et~al.}(2011){Jehin}, {Gillon}, {Queloz}, {Magain},
  {Manfroid}, {Chantry}, {Lendl}, {Hutsem{\'e}kers}, \& {Udry}}]{Jehin2011}
{Jehin}, E., {Gillon}, M., {Queloz}, D., {et~al.} 2011, The Messenger, 145, 2

\bibitem[{{Jenkins} {et~al.}(2016){Jenkins}, {Twicken}, {McCauliff},
  {Campbell}, {Sanderfer}, {Lung}, {Mansouri-Samani}, {Girouard}, {Tenenbaum},
  {Klaus}, {Smith}, {Caldwell}, {Chacon}, {Henze}, {Heiges}, {Latham},
  {Morgan}, {Swade}, {Rinehart}, \& {Vanderspek}}]{jenkins2016}
{Jenkins}, J.~M., {Twicken}, J.~D., {McCauliff}, S., {et~al.} 2016, in Society
  of Photo-Optical Instrumentation Engineers (SPIE) Conference Series, Vol.
  9913, Software and Cyberinfrastructure for Astronomy IV, 99133E

\bibitem[{Jones {et~al.}(2001)Jones, Oliphant, Peterson, {et~al.}}]{scipy}
Jones, E., Oliphant, T., Peterson, P., {et~al.} 2001, {SciPy}: Open source
  scientific tools for {Python}

\bibitem[{Kajtazi {et~al.}(2023)Kajtazi, Petit, \& Johansen}]{Kajtazi2023}
Kajtazi, K., Petit, A.~C., \& Johansen, A. 2023, Astronomy \& Astrophysics,
  669, A44

\bibitem[{Kass \& Raftery(1995)}]{kass1995}
Kass, R.~E. \& Raftery, A.~E. 1995, Journal of the American Statistical
  Association, 90, 773

\bibitem[{{Kempton} {et~al.}(2018){Kempton}, {Bean}, {Louie}, {Deming}, {Koll},
  {Mansfield}, {Christiansen}, {L{\'o}pez-Morales}, {Swain}, {Zellem},
  {Ballard}, {Barclay}, {Barstow}, {Batalha}, {Beatty}, {Berta-Thompson},
  {Birkby}, {Buchhave}, {Charbonneau}, {Cowan}, {Crossfield}, {de Val-Borro},
  {Doyon}, {Dragomir}, {Gaidos}, {Heng}, {Hu}, {Kane}, {Kreidberg}, {Mallonn},
  {Morley}, {Narita}, {Nascimbeni}, {Pall{\'e}}, {Quintana}, {Rauscher},
  {Seager}, {Shkolnik}, {Sing}, {Sozzetti}, {Stassun}, {Valenti}, \& {von
  Essen}}]{kempton2018}
{Kempton}, E. M.~R., {Bean}, J.~L., {Louie}, D.~R., {et~al.} 2018, \pasp, 130,
  114401

\bibitem[{{Kipping}(2013)}]{kipping2013}
{Kipping}, D.~M. 2013, \mnras, 435, 2152

\bibitem[{{Koornneef} {et~al.}(1986){Koornneef}, {Bohlin}, {Buser}, {Horne}, \&
  {Turnshek}}]{koornneef1986}
{Koornneef}, J., {Bohlin}, R., {Buser}, R., {Horne}, K., \& {Turnshek}, D.
  1986, Highlights of Astronomy, 7, 833

\bibitem[{{Kreidberg}(2015)}]{kreidberg2015}
{Kreidberg}, L. 2015, \pasp, 127, 1161

\bibitem[{{Kubyshkina} {et~al.}(2020){Kubyshkina}, {Vidotto}, {Fossati}, \&
  {Farrell}}]{Kubyshkina20}
{Kubyshkina}, D., {Vidotto}, A.~A., {Fossati}, L., \& {Farrell}, E. 2020,
  \mnras, 499, 77

\bibitem[{{Linder} {et~al.}(2019){Linder}, {Mordasini}, {Molli{\`e}re},
  {Marleau}, {Malik}, {Quanz}, \& {Meyer}}]{Linder19}
{Linder}, E.~F., {Mordasini}, C., {Molli{\`e}re}, P., {et~al.} 2019, \aap, 623,
  A85

\bibitem[{{Lithwick} {et~al.}(2012){Lithwick}, {Xie}, \& {Wu}}]{lithwick2012}
{Lithwick}, Y., {Xie}, J., \& {Wu}, Y. 2012, \apj, 761, 122

\bibitem[{{Lopez} {et~al.}(2012){Lopez}, {Fortney}, \& {Miller}}]{Lopez12}
{Lopez}, E.~D., {Fortney}, J.~J., \& {Miller}, N. 2012, \apj, 761, 59

\bibitem[{{Mann} {et~al.}(2016){Mann}, {Newton}, {Rizzuto}, {Irwin}, {Feiden},
  {Gaidos}, {Mace}, {Kraus}, {James}, {Ansdell}, {Charbonneau}, {Covey},
  {Ireland}, {Jaffe}, {Johnson}, {Kidder}, \& {Vanderburg}}]{mann2016}
{Mann}, A.~W., {Newton}, E.~R., {Rizzuto}, A.~C., {et~al.} 2016, \aj, 152, 61

\bibitem[{{Mann} {et~al.}(2022){Mann}, {Wood}, {Schmidt}, {Barber}, {Owen},
  {Tofflemire}, {Newton}, {Mamajek}, {Bush}, {Mace}, {Kraus}, {Thao},
  {Vanderburg}, {Llama}, {Johns-Krull}, {Prato}, {Stahl}, {Tang}, {Fields},
  {Collins}, {Collins}, {Gan}, {Jensen}, {Kamler}, {Schwarz}, {Furlan},
  {Gnilka}, {Howell}, {Lester}, {Owens}, {Suarez}, {Mekarnia}, {Guillot},
  {Abe}, {Triaud}, {Johnson}, {Milburn}, {Rizzuto}, {Quinn}, {Kerr}, {Ricker},
  {Vanderspek}, {Latham}, {Seager}, {Winn}, {Jenkins}, {Guerrero}, {Shporer},
  {Schlieder}, {McLean}, \& {Wohler}}]{mann2022}
{Mann}, A.~W., {Wood}, M.~L., {Schmidt}, S.~P., {et~al.} 2022, \aj, 163, 156

\bibitem[{{Martioli} {et~al.}(2021){Martioli}, {H{\'e}brard}, {Correia},
  {Laskar}, \& {Lecavelier des Etangs}}]{martioli2021}
{Martioli}, E., {H{\'e}brard}, G., {Correia}, A.~C.~M., {Laskar}, J., \&
  {Lecavelier des Etangs}, A. 2021, \aap, 649, A177

\bibitem[{{Matsumura} {et~al.}(2010){Matsumura}, {Peale}, \&
  {Rasio}}]{Matsumura10}
{Matsumura}, S., {Peale}, S.~J., \& {Rasio}, F.~A. 2010, \apj, 725, 1995

\bibitem[{{M{\'e}karnia} {et~al.}(2016){M{\'e}karnia}, {Guillot}, {Rivet},
  {Schmider}, {Abe}, {Gon{\c{c}}alves}, {Agabi}, {Crouzet}, {Fruth},
  {Barbieri}, {Bayliss}, {Zhou}, {Aristidi}, {Szulagyi}, {Daban},
  {Fante{\"\i}-Caujolle}, {Gouvret}, {Erikson}, {Rauer}, {Bouchy}, {Gerakis},
  \& {Bouchez}}]{Mekarnia2016}
{M{\'e}karnia}, D., {Guillot}, T., {Rivet}, J.~P., {et~al.} 2016, \mnras, 463,
  45

\bibitem[{{Mordasini}(2020)}]{Mordasini20}
{Mordasini}, C. 2020, \aap, 638, A52

\bibitem[{{Mordasini} {et~al.}(2012){Mordasini}, {Alibert}, {Georgy},
  {Dittkrist}, {Klahr}, \& {Henning}}]{Mordasini12}
{Mordasini}, C., {Alibert}, Y., {Georgy}, C., {et~al.} 2012, \aap, 547, A112

\bibitem[{{M{\"u}ller} {et~al.}(2020){M{\"u}ller}, {Ben-Yami}, \&
  {Helled}}]{Muller20}
{M{\"u}ller}, S., {Ben-Yami}, M., \& {Helled}, R. 2020, \apj, 903, 147

\bibitem[{{Nesvorn{\'y}} \& {Vokrouhlick{\'y}}(2016)}]{nesvorny2016}
{Nesvorn{\'y}}, D. \& {Vokrouhlick{\'y}}, D. 2016, \apj, 823, 72

\bibitem[{{Owen} \& {Wu}(2017)}]{Owen17}
{Owen}, J.~E. \& {Wu}, Y. 2017, \apj, 847, 29

\bibitem[{Papaloizou \& Terquem(2010)}]{Papaloizou2010}
Papaloizou, J. C.~B. \& Terquem, C. 2010, Monthly Notices of the Royal
  Astronomical Society, 405, 573

\bibitem[{{Plavchan} {et~al.}(2020){Plavchan}, {Barclay}, {Gagn{\'e}}, {Gao},
  {Cale}, {Matzko}, {Dragomir}, {Quinn}, {Feliz}, {Stassun}, {Crossfield},
  {Berardo}, {Latham}, {Tieu}, {Anglada-Escud{\'e}}, {Ricker}, {Vanderspek},
  {Seager}, {Winn}, {Jenkins}, {Rinehart}, {Krishnamurthy}, {Dynes}, {Doty},
  {Adams}, {Afanasev}, {Beichman}, {Bottom}, {Bowler}, {Brinkworth}, {Brown},
  {Cancino}, {Ciardi}, {Clampin}, {Clark}, {Collins}, {Davison},
  {Foreman-Mackey}, {Furlan}, {Gaidos}, {Geneser}, {Giddens}, {Gilbert},
  {Hall}, {Hellier}, {Henry}, {Horner}, {Howard}, {Huang}, {Huber}, {Kane},
  {Kenworthy}, {Kielkopf}, {Kipping}, {Klenke}, {Kruse}, {Latouf}, {Lowrance},
  {Mennesson}, {Mengel}, {Mills}, {Morton}, {Narita}, {Newton}, {Nishimoto},
  {Okumura}, {Palle}, {Pepper}, {Quintana}, {Roberge}, {Roccatagliata},
  {Schlieder}, {Tanner}, {Teske}, {Tinney}, {Vanderburg}, {von Braun}, {Walp},
  {Wang}, {Wang}, {Weigand}, {White}, {Wittenmyer}, {Wright}, {Youngblood},
  {Zhang}, \& {Zilberman}}]{plavchan2020}
{Plavchan}, P., {Barclay}, T., {Gagn{\'e}}, J., {et~al.} 2020, \nat, 582, 497

\bibitem[{{Rackham} {et~al.}(2018){Rackham}, {Apai}, \&
  {Giampapa}}]{rackham2018}
{Rackham}, B.~V., {Apai}, D., \& {Giampapa}, M.~S. 2018, \apj, 853, 122

\bibitem[{{Rajpaul} {et~al.}(2015){Rajpaul}, {Aigrain}, {Osborne}, {Reece}, \&
  {Roberts}}]{rajpaul2015}
{Rajpaul}, V., {Aigrain}, S., {Osborne}, M.~A., {Reece}, S., \& {Roberts}, S.
  2015, \mnras, 452, 2269

\bibitem[{{Rappaport} {et~al.}(2013){Rappaport}, {Deck}, {Levine}, {Borkovits},
  {Carter}, {El Mellah}, {Sanchis-Ojeda}, \& {Kalomeni}}]{rappaport2013}
{Rappaport}, S., {Deck}, K., {Levine}, A., {et~al.} 2013, \apj, 768, 33

\bibitem[{Rasmussen \& Williams(2005)}]{rasmussenwilliams2005}
Rasmussen, C.~E. \& Williams, C. K.~I. 2005, Gaussian Processes for Machine
  Learning (Adaptive Computation and Machine Learning) (The MIT Press)

\bibitem[{{Rein} \& {Liu}(2012)}]{rein2012}
{Rein}, H. \& {Liu}, S.~F. 2012, \aap, 537, A128

\bibitem[{{Rein} \& {Spiegel}(2015)}]{rein2015}
{Rein}, H. \& {Spiegel}, D.~S. 2015, \mnras, 446, 1424

\bibitem[{{Ricker} {et~al.}(2015){Ricker}, {Winn}, {Vanderspek}, {Latham},
  {Bakos}, {Bean}, {Berta-Thompson}, {Brown}, {Buchhave}, {Butler}, {Butler},
  {Chaplin}, {Charbonneau}, {Christensen-Dalsgaard}, {Clampin}, {Deming},
  {Doty}, {De Lee}, {Dressing}, {Dunham}, {Endl}, {Fressin}, {Ge}, {Henning},
  {Holman}, {Howard}, {Ida}, {Jenkins}, {Jernigan}, {Johnson}, {Kaltenegger},
  {Kawai}, {Kjeldsen}, {Laughlin}, {Levine}, {Lin}, {Lissauer}, {MacQueen},
  {Marcy}, {McCullough}, {Morton}, {Narita}, {Paegert}, {Palle}, {Pepe},
  {Pepper}, {Quirrenbach}, {Rinehart}, {Sasselov}, {Sato}, {Seager},
  {Sozzetti}, {Stassun}, {Sullivan}, {Szentgyorgyi}, {Torres}, {Udry}, \&
  {Villasenor}}]{ricker2015}
{Ricker}, G.~R., {Winn}, J.~N., {Vanderspek}, R., {et~al.} 2015, Journal of
  Astronomical Telescopes, Instruments, and Systems, 1, 014003

\bibitem[{{Rodrigo} \& {Solano}(2020)}]{rodrigo2020}
{Rodrigo}, C. \& {Solano}, E. 2020, in XIV.0 Scientific Meeting (virtual) of
  the Spanish Astronomical Society, 182

\bibitem[{{Rodrigo} {et~al.}(2012){Rodrigo}, {Solano}, \& {Bayo}}]{rodrigo2012}
{Rodrigo}, C., {Solano}, E., \& {Bayo}, A. 2012, {SVO Filter Profile Service
  Version 1.0}, IVOA Working Draft 15 October 2012

\bibitem[{{Schmider} {et~al.}(2022){Schmider}, {Abe}, {Agabi}, {Bendjoya},
  {Crouzet}, {Dransfield}, {Guillot}, {Lai}, {Mekarnia}, {Suarez}, {Triaud},
  {Stee}, {G{\"u}nther}, {Breeveld}, \& {Blommaert}}]{Schmider2022}
{Schmider}, F.-X., {Abe}, L., {Agabi}, A., {et~al.} 2022, in Society of
  Photo-Optical Instrumentation Engineers (SPIE) Conference Series, Vol. 12182,
  Ground-based and Airborne Telescopes IX, ed. H.~K. {Marshall},
  J.~{Spyromilio}, \& T.~{Usuda}, 121822O

\bibitem[{{Sebastian} {et~al.}(2021){Sebastian}, {Gillon}, {Ducrot},
  {Pozuelos}, {Garcia}, {G{\"u}nther}, {Delrez}, {Queloz}, {Demory}, {Triaud},
  {Burgasser}, {de Wit}, {Burdanov}, {Dransfield}, {Jehin}, {McCormac},
  {Murray}, {Niraula}, {Pedersen}, {Rackham}, {Sohy}, {Thompson}, \& {Van
  Grootel}}]{Sebastian_2021AA}
{Sebastian}, D., {Gillon}, M., {Ducrot}, E., {et~al.} 2021, \aap, 645, A100

\bibitem[{{Skrutskie} {et~al.}(2006){Skrutskie}, {Cutri}, {Stiening},
  {Weinberg}, {Schneider}, {Carpenter}, {Beichman}, {Capps}, {Chester},
  {Elias}, {Huchra}, {Liebert}, {Lonsdale}, {Monet}, {Price}, {Seitzer},
  {Jarrett}, {Kirkpatrick}, {Gizis}, {Howard}, {Evans}, {Fowler}, {Fullmer},
  {Hurt}, {Light}, {Kopan}, {Marsh}, {McCallon}, {Tam}, {Van Dyk}, \&
  {Wheelock}}]{Skrutskie2006}
{Skrutskie}, M.~F., {Cutri}, R.~M., {Stiening}, R., {et~al.} 2006, \aj, 131,
  1163

\bibitem[{{Smith} {et~al.}(2012){Smith}, {Stumpe}, {Van Cleve}, {Jenkins},
  {Barclay}, {Fanelli}, {Girouard}, {Kolodziejczak}, {McCauliff}, {Morris}, \&
  {Twicken}}]{smith2012}
{Smith}, J.~C., {Stumpe}, M.~C., {Van Cleve}, J.~E., {et~al.} 2012, \pasp, 124,
  1000

\bibitem[{{Speagle}(2020)}]{speagle2020}
{Speagle}, J.~S. 2020, \mnras, 493, 3132

\bibitem[{{Stumpe} {et~al.}(2014){Stumpe}, {Smith}, {Catanzarite}, {Van Cleve},
  {Jenkins}, {Twicken}, \& {Girouard}}]{stumpe2014}
{Stumpe}, M.~C., {Smith}, J.~C., {Catanzarite}, J.~H., {et~al.} 2014, \pasp,
  126, 100

\bibitem[{{Stumpe} {et~al.}(2012){Stumpe}, {Smith}, {Van Cleve}, {Twicken},
  {Barclay}, {Fanelli}, {Girouard}, {Jenkins}, {Kolodziejczak}, {McCauliff}, \&
  {Morris}}]{stumpe2012}
{Stumpe}, M.~C., {Smith}, J.~C., {Van Cleve}, J.~E., {et~al.} 2012, \pasp, 124,
  985

\bibitem[{{Su{\'a}rez Mascare{\~n}o} {et~al.}(2021){Su{\'a}rez Mascare{\~n}o},
  {Damasso}, {Lodieu}, {Sozzetti}, {B{\'e}jar}, {Benatti}, {Zapatero Osorio},
  {Micela}, {Rebolo}, {Desidera}, {Murgas}, {Claudi}, {Gonz{\'a}lez
  Hern{\'a}ndez}, {Malavolta}, {del Burgo}, {D'Orazi}, {Amado}, {Locci},
  {Tabernero}, {Marzari}, {Aguado}, {Turrini}, {Cardona Guill{\'e}n},
  {Toledo-Padr{\'o}n}, {Maggio}, {Aceituno}, {Bauer}, {Caballero},
  {Chinchilla}, {Esparza-Borges}, {Gonz{\'a}lez-{\'A}lvarez}, {Granzer},
  {Luque}, {Mart{\'\i}n}, {Nowak}, {Oshagh}, {Pall{\'e}}, {Parviainen},
  {Quirrenbach}, {Reiners}, {Ribas}, {Strassmeier}, {Weber}, \&
  {Mallonn}}]{suarez2021}
{Su{\'a}rez Mascare{\~n}o}, A., {Damasso}, M., {Lodieu}, N., {et~al.} 2021,
  Nature Astronomy, 6, 232

\bibitem[{Teyssandier \& Libert(2020)}]{Teyssandier2020}
Teyssandier, J. \& Libert, A.-S. 2020, Astronomy and Astrophysics, 643, A11

\bibitem[{{Vazan} {et~al.}(2013){Vazan}, {Kovetz}, {Podolak}, \&
  {Helled}}]{Vazan13}
{Vazan}, A., {Kovetz}, A., {Podolak}, M., \& {Helled}, R. 2013, \mnras, 434,
  3283

\bibitem[{{Venturini} {et~al.}(2020){Venturini}, {Guilera}, {Haldemann},
  {Ronco}, \& {Mordasini}}]{Venturini20}
{Venturini}, J., {Guilera}, O.~M., {Haldemann}, J., {Ronco}, M.~P., \&
  {Mordasini}, C. 2020, \aap, 643, L1

\bibitem[{{Yu} {et~al.}(2015){Yu}, {Winn}, {Gillon}, {Albrecht}, {Rappaport},
  {Bieryla}, {Dai}, {Delrez}, {Hillenbrand}, {Holman}, {Howard}, {Huang},
  {Isaacson}, {Jehin}, {Lendl}, {Montet}, {Muirhead}, {Sanchis-Ojeda}, \&
  {Triaud}}]{yu2015}
{Yu}, L., {Winn}, J.~N., {Gillon}, M., {et~al.} 2015, \apj, 812, 48

\end{thebibliography}

\begin{appendix} 

\FloatBarrier
\section{Additional figures and tables}

\begin{figure}[h]
   \centering
   \includegraphics[width=0.49\textwidth]{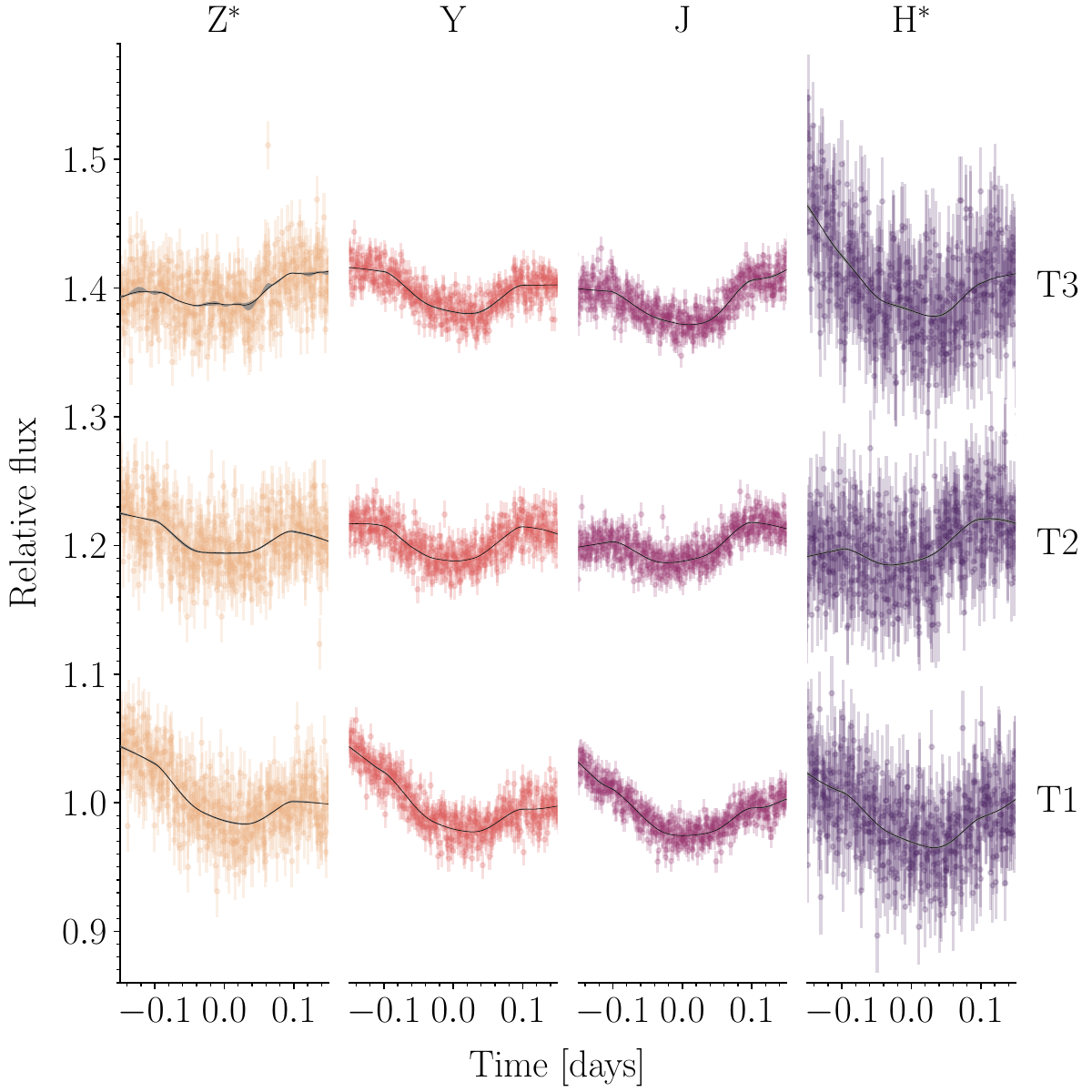}
      \caption{ExTrA full transit photometry observation of TOI-1227\,b in Z*, Y, J, and H* bands (error bars). Each line corresponds to an ExTrA telescope (labeled T1, T2, and T3) observation that is offset vertically for clarity. For each transit the median model (black line) and 68\% CI (gray band, barely visible) computed from 1000 random posterior samples are shown.}
      \label{fig.ExTrA_ZYJH}
\end{figure}

\begin{table}
  \setlength{\tabcolsep}{5pt}
\renewcommand{\arraystretch}{1.25}
\centering
\caption{Inferred $R_p/R_\star$ in different bands and TESS sectors.}\label{table.RpRsRatio}
\begin{tabular}{lrcc}
\hline
\hline
Band & $\lambda_{\rm pivot}$~[nm] & Prior & Median and 68.3\% CI \\
\hline
g'             &   476  & $U(0, 1)$ & $0.142 \pm 0.011$ \\
TESS s11       &   770  & $U(0, 1)$ & $0.176 \pm 0.018$ \\
TESS s12       &   770  & $U(0, 1)$ & $0.184 \pm 0.024$ \\
TESS s38       &   770  & $U(0, 1)$ & $0.161 \pm 0.014$ \\
TESS s64       &   770  & $U(0, 1)$ & $0.152 \pm 0.015$ \\
TESS s65       &   770  & $U(0, 1)$ & $0.152 \pm 0.013$ \\
i'             &   772  & $U(0, 1)$ & $0.1463 \pm 0.0087$ \\
ASTEP+ Red     &   819  & $U(0, 1)$ & $0.1430 \pm 0.0093$ \\
\textit{I+z}   &   850  & $U(0, 1)$ & $0.1628 \pm 0.0075$ \\
z$_{\rm S}$    &   867  & $U(0, 1)$ & $0.1549_{-0.0048}^{+0.0056}$ \\
ExTrA $Z^{*}$  &   890  & $U(0, 1)$ & $0.154 \pm 0.015$ \\
z'             &$\sim$912  & $U(0, 1)$ & $0.1581 \pm 0.0062$ \\
ExTrA $Y$      &  1031  & $U(0, 1)$ & $0.160 \pm 0.011$ \\
ExTrA          &  1165  & $U(0, 1)$ & $0.163 \pm 0.017$ \\
ExTrA $J$      &  1248  & $U(0, 1)$ & $0.165 \pm 0.016$ \\
ExTrA $H{*}$   &  1526  & $U(0, 1)$ & $0.151_{-0.028}^{+0.025}$ \\
\hline
\end{tabular}
\tablefoot{The table lists: band, $\lambda_{\rm pivot}$ \citep{koornneef1986}, prior, posterior median, and 68.3\% CI. $U(a, b)$: A uniform distribution defined between a lower $a$ and an upper $b$ limit.}
\end{table}

\begin{figure}[h]
   \centering
   \includegraphics[width=0.49\textwidth]{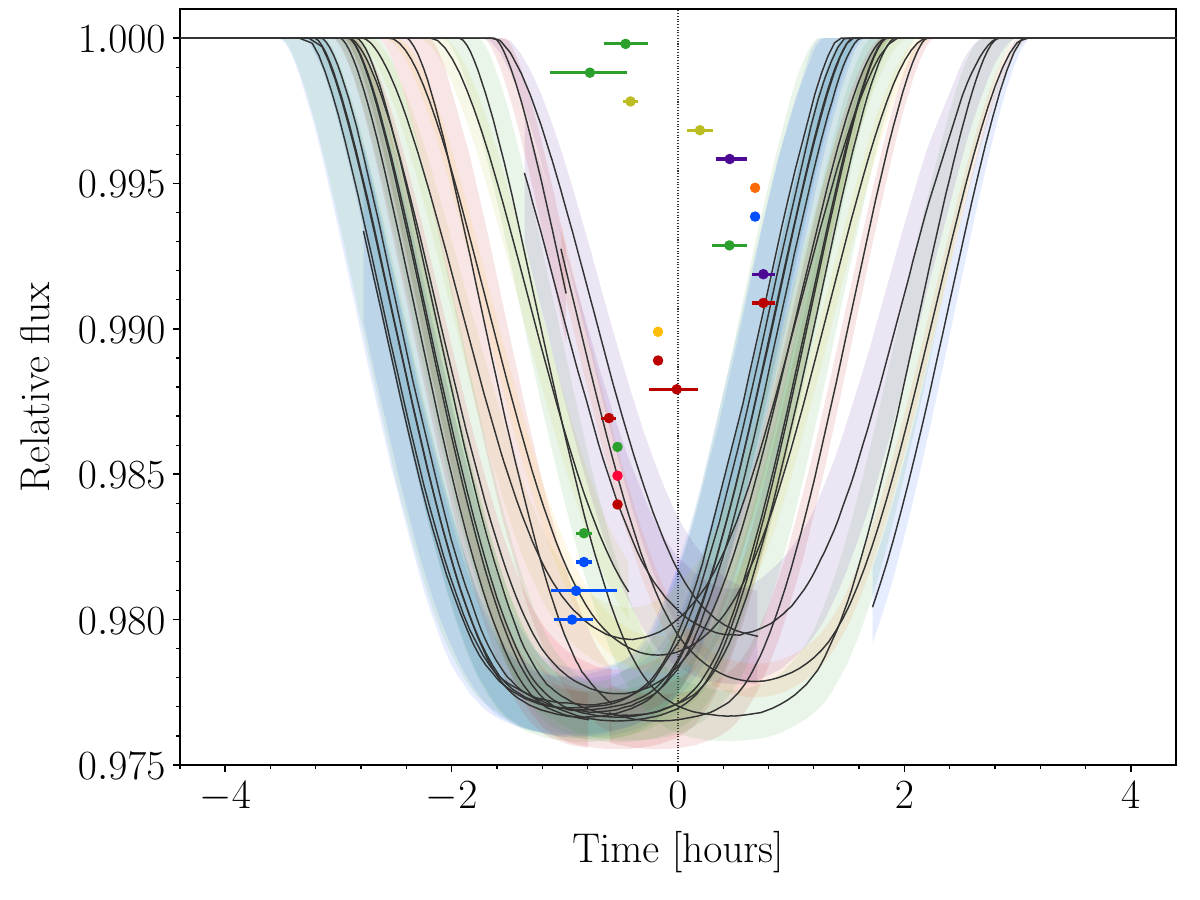}
      \caption{Posterior of the transit models for each epoch and band (Sect.~\ref{section.juliet}, TTV analysis with a single $R_p/R_\star$), with the median represented as a black line and the 68\% credible interval shown in various colors corresponding to Fig.~\ref{fig.phot}, excludes epoch 16. Each transit is centered relative to the MAP value of the linear ephemeris derived in Sect. 4.1. The horizontal error bars indicate the mid-transit time offset from the linear ephemeris, shifted vertically, with time increasing from top to bottom.}
      \label{fig.model}
\end{figure}

\section{Supplementary assessment}\label{sec.referee}

\begin{figure}[h]
   \centering
   \includegraphics[width=0.49\textwidth]{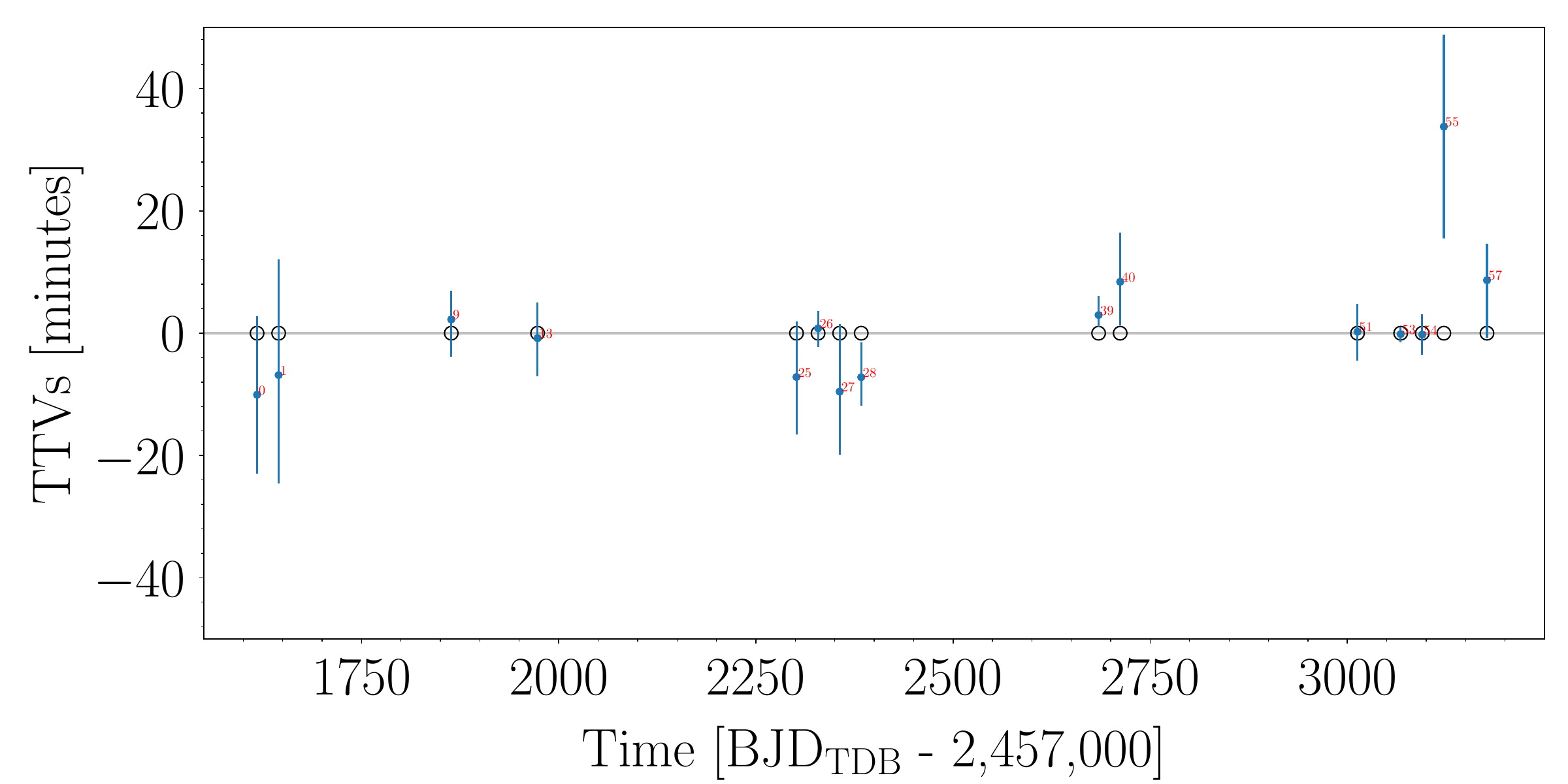}
   \includegraphics[width=0.49\textwidth]{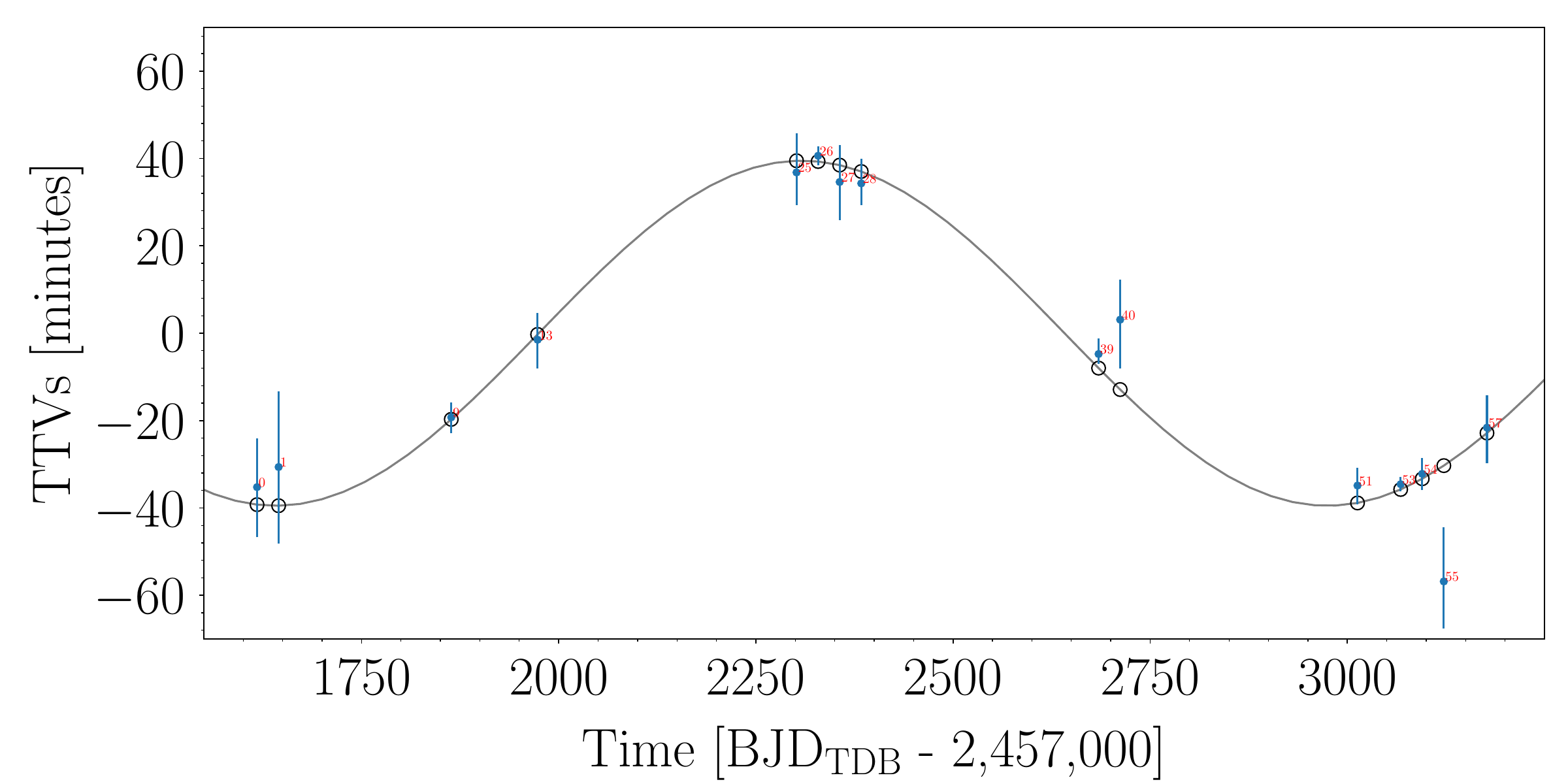}
      \caption{Posterior TTVs of the simulated data (simulation 1 in the upper panel and simulation 2 in the lower panel), relative to a linear ephemeris, are shown with blue error bars. The open black circles are the simulated data, and the gray line is the simulated model for the TTVs.}
      \label{fig.sim}
\end{figure}

In order to validate the accuracy of our methodology in determining transit times, we applied the methods outlined in Sect.~\ref{section.juliet} to simulated data. We started from the true observation, removed the MAP model from the analysis presented Sect.~\ref{section.juliet}, and then added back the same model but with a different mid-transit time. The simulated times were pure linear ephemeris (no TTVs) for simulation 1, and the MAP sinusoidal model of Sect.~\ref{section.sine} for simulation 2. This procedure tests the original data (with its systematics and sampling) at slightly different times. We have excluded epoch 16, which was not used for Sect.~\ref{section.model}. Figure~\ref{fig.sim} compares posterior and simulated transit times. To further compare the posteriors with the simulated values, we calculated the offset normalized by the 68.3\% CI. This metric has a mean of 0.11 and 0.20, and a standard deviation of 0.88 and 0.86, for simulations 1 and 2, respectively. This is compatible with a standard normal distribution, given the small number of samples, suggesting that the systematics do not prevent an accurate estimation of the transit timing.

\end{appendix}
\end{document}